\documentclass[journal,twoside,web]{ieeecolor}

\usepackage{tikz}
\usepackage{textcomp}
\usepackage{hyperref}
\usepackage{lipsum}

\newcommand\copyrighttext{%
  \footnotesize \textcopyright 2022 IEEE. Personal use of this material is permitted.
  Permission from IEEE must be obtained for all other uses, in any current or future
  media, including reprinting/republishing this material for advertising or promotional
  purposes, creating new collective works, for resale or redistribution to servers or
  lists, or reuse of any copyrighted component of this work in other works.
  DOI: \href{https://doi.org/10.1109/TMI.2022.3203309}{10.1109/TMI.2022.3203309}}
\newcommand\copyrightnotice{%
\begin{tikzpicture}[remember picture,overlay]
\node[anchor=north,yshift=-150pt] at (current page.north) {\fbox{\parbox{\dimexpr\textwidth-\fboxsep-\fboxrule\relax}{\copyrighttext}}};
\end{tikzpicture}%
}

\newcommand{\markupcolor}{black} 
\usepackage[disable,color=white,textsize=small,textcolor=\markupcolor,bordercolor=\markupcolor,textwidth=1cm]{todonotes} 
\setuptodonotes{noline}
\usepackage{marginnote}

\usepackage{tmi}
\usepackage{cite}
\usepackage{amsmath,amssymb,amsfonts}
\usepackage{algorithmic}
\usepackage{graphicx}
\usepackage{textcomp}

\usepackage{hyperref}
\usepackage{balance}
\usepackage{xcolor}
\let\labelindent\relax\usepackage{enumitem}
\usepackage{tabularx}
\usepackage{colortbl}
\usepackage{booktabs}
\newcommand{\iqr}{5pt}
\newcolumntype{C}{>{\centering\arraybackslash}X}
\newcolumntype{L}{>{\raggedright\arraybackslash}X}
\newcolumntype{R}{>{\raggedleft\arraybackslash}X}
\usepackage{scalerel}
\usepackage{stfloats}
\DeclareRobustCommand{\markup}[1]{{\color{\markupcolor}{#1}}}

\def\BibTeX{{\rm B\kern-.05em{\sc i\kern-.025em b}\kern-.08em
    T\kern-.1667em\lower.7ex\hbox{E}\kern-.125emX}}
\begin{document}
\title{A persistent homology-based topological loss for CNN-based multi-class segmentation of CMR}
\author{Nick Byrne, James R. Clough, Israel Valverde, Giovanni Montana, Andrew P. King
\thanks{Manuscript received 30 July 2021; revised 3 July 2022; accepted 23 August 2022.
Nick Byrne is funded by a National Institute for Health Research (NIHR), Doctoral Research Fellowship for this research project. This report presents independent research funded by the NIHR. The views expressed are those of the author(s) and not necessarily those of the NHS, the NIHR or the Department of Health and Social Care.
This work was supported by the EPSRC programme Grant ‘SmartHeart’ (EP/P001009/1), the Wellcome Trust IEH Award [102431], the Wellcome EPSRC Centre for Medical Engineering at School of Biomedical Engineering and Imaging Sciences (BMEIS), King's College London (KCL) (WT 203148/Z/16/Z) and by the NIHR Biomedical Research Centre at Guys and St Thomas NHS Foundation Trust (GSTT) and King's College London.}
\thanks{Nick Byrne (nicholas.byrne@kcl.ac.uk), James R. Clough (james.clough91@gmail.com), Isra Valverde (isra.valverde@kcl.ac.uk) and Andrew P. King (andrew.king@kcl.ac.uk) are with the School of BMEIS at KCL: London, SE1 7EH, UK.
Nick Byrne and Isra Valverde are also with the Medical Physics and Paediatric Cardiology Departments respectively, both at GSTT: London, SE1 7EH, UK.
Giovanni Montana (g.montana@warwick.ac.uk) is with the Warwick Manufacturing Group at the University of Warwick: Coventry, CV4 7AL, UK.}
}

\maketitle
\copyrightnotice
\vspace{-3.5mm}

\begin{abstract}
Multi-class segmentation of cardiac magnetic resonance (CMR) images seeks a separation of data into anatomical components with known structure and configuration.
The most popular CNN-based methods are optimised using pixel wise loss functions, ignorant of the spatially extended features that characterise anatomy.
Therefore, whilst sharing a high spatial overlap with the ground truth, inferred CNN-based segmentations can lack coherence, including spurious connected components, holes and voids.
Such results are implausible, violating anticipated anatomical topology.
In response, (single-class) persistent homology-based loss functions have been proposed to capture global anatomical features.
Our work extends these approaches to the task of multi-class segmentation.
Building an enriched topological description of all class labels and class label pairs, our loss functions make predictable and statistically significant improvements in segmentation topology using a CNN-based post-processing framework.
We also present (and make available) a highly efficient implementation based on cubical complexes and parallel execution, enabling practical application within high resolution 3D data for the first time.
We demonstrate our approach on 2D short axis and 3D whole heart CMR segmentation, advancing a detailed and faithful analysis of performance on two publicly available datasets.
\end{abstract}

\begin{IEEEkeywords}
CMR, CNN, Image segmentation, Topology
\end{IEEEkeywords}
\section{Introduction}
\label{sec:intro}

\IEEEPARstart{M}{edical} image segmentation is a prerequisite of many pipelines dedicated to the analysis and visualisation of clinical data.
A pervasive example from cardiac magnetic resonance (CMR) divides the 2D short axis image into myocardial, and ventricular components.
Associated segmentations permit quantitative assessment of important clinical indices \cite{Ruijsink2019} and the analysis of cardiac physiology \cite{LopezPerez2015}.
Further to their quantitative application, segmented CMR images find a qualitative role within the visualisation of 3D data, including the planning of structural intervention \cite{Valverde2017}.
These successes have been achieved in spite of the operator burden of manual image segmentation, a task that demands specialist expertise and which in 3D, can take hours to complete \cite{Byrne2016}.

In response, deep learning, and in particular convolutional neural networks (CNNs), have fostered significant gains \cite{Chen2020}.
Key to their success has been the design of specialised architectures for image segmentation, U-Net \cite{Ronneberger2015} being the prime example.
Implicit within this multi-scale architecture is an acknowledgement that image segments are discriminated by pixel, local and global image features, including anatomical morphology.
Whilst considerable effort has been devoted to the extraction of multi-scale image features, less attention has been paid to their role in optimisation \cite{Duan2019}.
For the most part, CNNs have been trained using pixel-wise loss functions such as cross-entropy (CE) or the Dice similarity coefficient (DSC).
Whilst easily implemented and possessing favourable numerical properties, their treatment of pixels as independent from one another renders them insensitive to higher order features of anatomy.
This is in contrast with the spatially extended features that we anticipate are learned during training.

Without explicitly considering such features, CNN optimisation can predict segmentations which lack spatial coherence, including spurious connected components or holes \cite{Painchaud2019}.
To the operator, such errors appear nonsensical, violating fundamental properties of anatomy.
Since they are often small and constrained to object boundaries, their associated segmentation may remain suitable for the assessment of certain clinical indices.
However, for a wider array of downstream applications it is crucial to represent such features faithfully \cite{Byrne2016}.

Historically, and before the advent of deep learning, state of the art cardiac segmentation methods depended on strong priors describing anatomical configuration,
such as atlas-based approaches \cite{Zuluaga2013} or those based on deformable or statistical shape models \cite{Wierzbicki2008}.
Ideally, however, such prior information is abstract and adaptable to the variety of cases encountered in the clinic, not
dependent on its appearance within exemplar training data.
In the simplest case, abstract priors might specify the configuration of the heart's chambers, valves and associated vasculature.
For example, in short axis CMR the right ventricular cavity is bound to the left ventricular myocardium, which in turn surrounds the left ventricular blood pool.

Whether describing normal or pathological anatomy, and in contrast to the pixel-wise losses mentioned, anatomical priors provide a global description of segmentation coherence.
However, whilst these constraints are simple to express qualitatively, their quantitative expression is not trivial.
Furthermore, the opacity of CNNs has hindered efforts to explicitly utilise such priors in parameter optimisation.
Addressing these limitations, we present a topological loss function for multi-class image segmentation, leveraging an explicit anatomical description that is independent of exemplar training data.

\section{Previous work}
\label{sec:lit_rev}

\begin{figure*}[t]
\centering
\hspace*{\fill}%
\minipage{0.2\linewidth}
  \centering
  \includegraphics[width=0.9\linewidth]{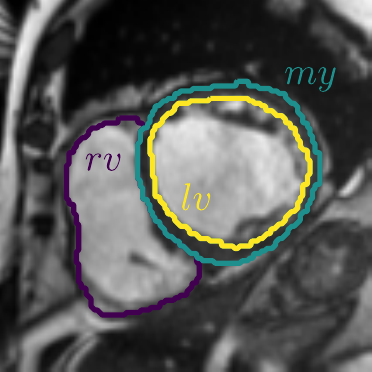}
  (a) 2D Short axis
\endminipage
\hfill
\minipage{0.62\linewidth}
  \centering
  \includegraphics[width=0.9\linewidth]{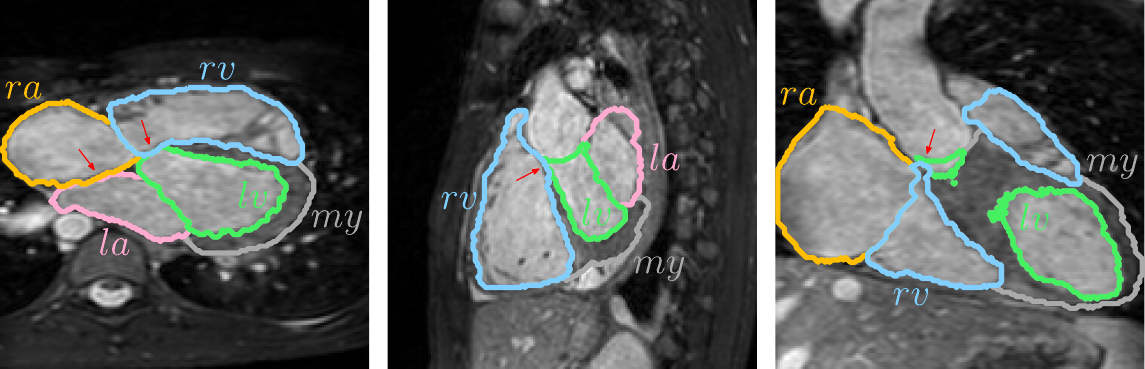}\todo{R3.7a} \\
  (b) 3D Whole heart
\endminipage
\hspace*{\fill}
\caption{\markup{(a) Short axis segmentation of 2D ACDC data \cite{Bernard2018} into right ($rv$) and left ($lv$) ventricular cavities, and myocardium ($my$).
(b) Whole heart segmentation of 3D MM-WHS data \cite{Zhuang2019} into right ($ra$) and left ($la$) atria, in addition to $rv$, $lv$ and $my$.
Often associated with anomalous connections between the left and right heart, arrows indicate errors in multi-class label map topology that, whilst present within provided data, are consistent with the challenge's focus on spatial overlap.
Demanding attention to pixel-level adjacency, they are likely artefacts of manual segmentation.
}}\label{fig:tasks}
\end{figure*}

\subsection{Anatomical priors within cardiac segmentation}
\label{subsec:cardiac_lit}

CNNs have emerged as the state of the art reference standard for cardiac image segmentation, achieving notable success within: 2D short axis CMR segmentation \cite{Bernard2018}; and 3D CT/CMR multi-class labelling \cite{Zhuang2019}.
We consider closely related tasks in the remainder of this paper (see \autoref{fig:tasks}).

\subsubsection{2D short axis segmentation}

Short axis segmentation has received the greatest interest of any cardiac task \cite{Chen2020}.
In studies of healthy subjects, this has culminated in a level of performance consistent with interobserver variation \cite{Bai2018}.
However, in studies of cardiovascular disease, a performance deficit remains.
This gap is partially characterised by the type of anatomically implausible error described in \autoref{sec:intro}.

To address these modes of failure, authors have sought to leverage prior information.
One approach has combined CNNs with conventional methods such as atlas-based segmentation \cite{Duan2019}, active contours \cite{Ngo2017} or statistical shape models \cite{Tothova2018}.
Other works have attempted to inject prior information into CNN optimisation directly, via a learned, latent representation of anatomically plausible shapes \cite{Oktay2018}.
These implicit embeddings, however, make it difficult to understand the extent to which such priors are related to morphology or topology as claimed.
Work in \cite{Painchaud2019} bridged this gap, augmenting the latent space by only maintaining anatomically plausible cases.

\subsubsection{3D whole heart segmentation}

Before the shift to deep learning, whole heart segmentation made use of strong prior knowledge, with
statistical shape models \cite{Wierzbicki2008} and in particular atlas-based segmentation \cite{Zuluaga2013} representing the state of the art.
In contrast, exponents of CNN-based segmentation have been necessarily preoccupied with solutions for handling large 3D volumes.
Work has focused on architectural modifications including cascaded spatial processing, patch-based inference and slice-by-slice or 2.5D segmentation \cite{Isensee2018}.

There are exceptions to this trend, however: \cite{Wang2017} incorporated the results of statistical shape modelling as an additional CNN input channel; \cite{XuXiaowei2020} used a graphical representation of the great vessels to improve diagnostic classification and segmentation.
Works combining multi atlas-based registration and CNNs have also been presented \cite{Luo2020a}.
Most recently, \cite{Wang2021} employed a latent representation of cardiac anatomy within a few shot learning framework, achieving impressive results.

\subsection{Segmentation topology}
\label{subsec:wider_lit}

In other clinical settings, priors specifying the adjacency \cite{Ganaye2018}, containment \cite{BenTaieb2016} and hierarchy \cite{He2019} of anatomical components have been used to build associated loss functions.
We observe that across applications \cite{BenTaieb2016,Ganaye2018,Painchaud2019}, criteria defining anatomical plausibility can frequently be summarised by label map topology.
Recently, persistent homology (PH), an emerging tool for topological data analysis, has been combined with machine learning \cite{Hofer2017,Bruel-Gabrielsson2019}.
Associated topological features have been employed within image classification and segmentation by $k$-means clustering \cite{Assaf2016} and $k$-nearest neighbour \cite{Qaiser2016} search.
They have also been leveraged to distill dermoscopic images \cite{Vandaele2020}.
Bespoke input layers have been used to feed PH features to CNNs, improving electroencephalogram classification \cite{Hofer2019}, and the detection of COVID-19 \cite{Hajij2021}.

Pertinently, PH has been employed not only as a source of features for classification, but also to extract a supervisory signal for optimisation; including within binary segmentation tasks.
Reference \cite{Hu2019} established a PH-based loss according to the Wasserstein distance, applying their approach to microscopy and within the natural image domain.
Other exponents of PH-based losses have optimised segmentation topology against an explicit prior.
Works have considered the topologies of the branching murine neurovasculature \cite{Haft-Javaherian2020}, the cylindrical small bowel \cite{Shin2020} and the toroidal appearance of the myocardium in short axis CMR \cite{Clough2019,Clough2020}.

\section{Contributions}

\markup{Compared with the binary case, multi-class treatment admits consideration of a richer set of topological priors, including}
\todo{AE.1 \\ R2.1 \\ R3.1}\markup{hierarchical class containment and adjacency.
Together with its preceding conference submission \cite{Byrne2021}, this work reports our extension of PH-based topological losses to multi-class, CNN-based segmentation, making the following contributions:}
\begin{enumerate}[before=\color{\markupcolor},nolistsep]
    \item In Reference \cite{Byrne2021} we established a novel combinatorial loss formulation to optimise multi-class topology.
    \item Here, we present and share a new, efficient implementation, based on cubical complexes and parallelisation.
    \item We leverage these gains within a post-processing framework, demonstrating significant improvements in topology on two CMR segmentation tasks:
        \begin{enumerate}
            \item Compared with Reference \cite{Byrne2021}, we reduce computation time, achieving a fifty fold acceleration without sacrificing performance on 2D short axis segmentation.
            \item In a four-chambered segmentation of isotropically high resolution data, we are the first to apply PH-based post-processing and our multi-class loss to 3D anatomy.
        \end{enumerate}
    \item Lastly, we present a faithful analysis of the formalism set out in Reference \cite{Byrne2021}, including its limitations when applied to a clinical task posing genuine challenge.
\end{enumerate}%

\section{Theory}
\label{sec:theory}

\subsection{Betti numbers, homology and cubical complexes}
\label{subsec:hom_cub_comp}

\begin{figure*}[t]
\centering
\includegraphics[width=0.85\linewidth]{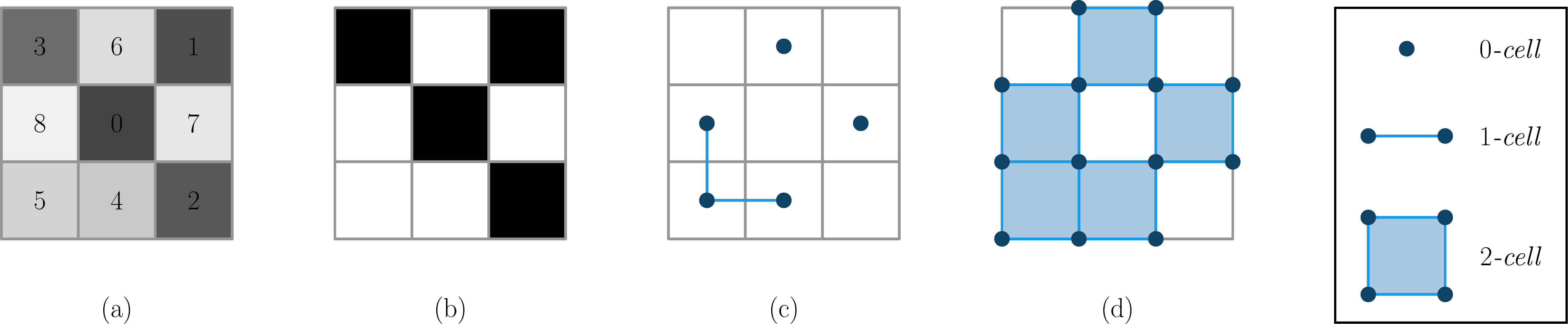}
\caption{
Construction of a cubical complex from 2D data.
Pixel intensities in (a) exceeding an arbitrary threshold of three appear white in the binary image (b).
In (c), these white pixels are considered 0-\emph{cells}, representing a 4-\emph{connected} foreground, including three components.
In (d) these are considered 2-\emph{cells}, representing an 8-\emph{connected} foreground component, containing a hole.
}
\label{fig:cub_comp}
\end{figure*}

In $N$D, objects with differing topology can be distinguished by the first $N$ Betti numbers: $\mathbf{b} = (b_0, b_1)$ in 2D and $\mathbf{b} = (b_0, b_1, b_2)$ in 3D.
Intuitively, $b_0$ counts the connected components which make up an object, $b_1$, the number of 2D holes or loops and $b_2$, the number of 3D holes or voids contained \cite{Otter2017}.
Whilst this characterisation is simple to express in abstraction, its practical computation relies upon homology, a branch of algebraic topology concerned with procedures to compute the topological features of objects.
The mathematical machinery underlying homology depends on the representation of objects by combinatorial structures called simplicial and cubical complexes \cite{Otter2017}.
Provided its representation as a valid complex, homology returns the Betti numbers of a binary segmentation by linear algebra alone.

Cubical complexes are well suited to data structured on a rectangular lattice, constructing 3D objects as the combination of points (0-\emph{cells}), and unit line intervals (1-\emph{cells}), squares (2-\emph{cells}) and cubes (3-\emph{cells}).
For image data, this representation is not only more computationally efficient and elegant than its simplicial equivalent (used in \cite{Bruel-Gabrielsson2019,Byrne2021,Clough2019}), but also affords control over the connectivity of pixels.
As illustrated by \autoref{fig:cub_comp}, there are two approaches for the construction of cubical complexes from 2D data \cite{Garin2020}.
These differ in their treatment of pixels as either $0$\emph{-cells} or $2$\emph{-cells} of the resulting complex, representing 4-\emph{connected} and 8-\emph{connected} objects, respectively.
Despite this difference, in initial experiments \markup{(see the Appendix in \autoref{sec:appendix})}\todo{NB.1a} we found an equivalence between topological loss functions when applied in conjunction with either mode of construction.
Accordingly, we consider foreground image pixels to be 4-\emph{connected} (or 6-\emph{connected} in 3D), and as 0-\emph{cells} in the remainder of this work.

\subsection{Persistent homology (PH)}

\begin{figure}[t]
\centering
\includegraphics[width=\linewidth]{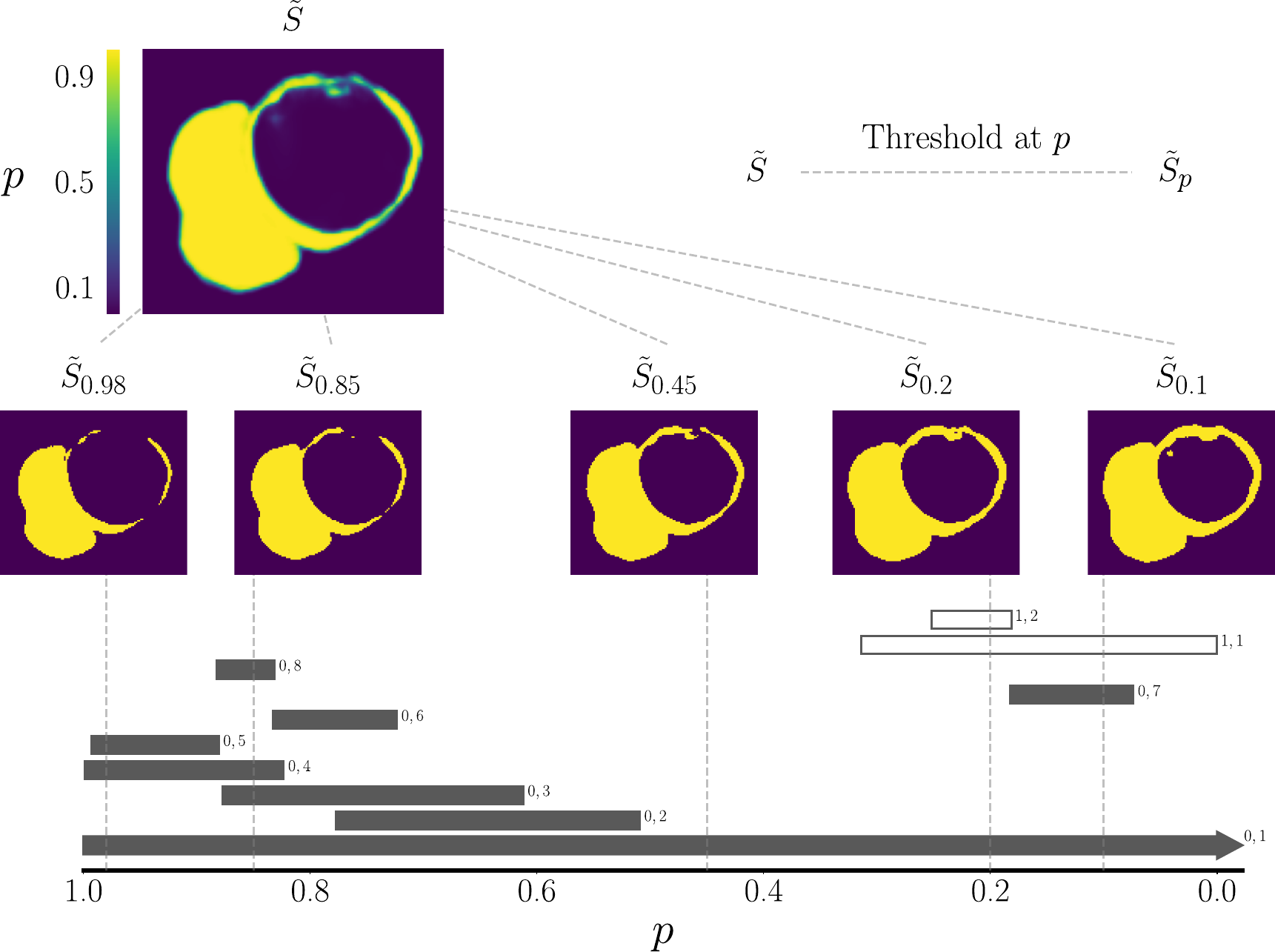}\todo{R3.9}
\caption{\markup{At a given value of the threshold, $p$, the PH barcode summarises the topological features presented by a probabilistic segmentation, $\tilde{S}$.
The number of vertically arranged solid bars count the connected components (0D holes); open bars count the loops (1D holes).
The width of each bar is the persistence ($\Delta p$) of each feature, between the critical values defining its birth and death.
Bars are labelled by topological dimension, and ranked in order of descending persistence: $d, l$.}}\label{fig:PH_barcode}
\end{figure}
In general, PH describes a scheme for the particular application of algebraic topology which exposes the topological features of data (their Betti numbers) at multiple scales \cite{Otter2017}.
We use the 2D example of \autoref{fig:PH_barcode} to illustrate the associated persistence barcode, providing a practical explanation as it relates to our application.
This considers the probabilistic segmentation $\tilde{S}: \mathbb{R}^{2} \rightarrow [ 0, 1 ]$.
Critically, rather than considering topology at a single probability threshold $\tilde{S}_{p} = \tilde{S} \ge p$, PH captures the topology of $\tilde{S}_{p}$ at \emph{all} possible thresholds.
Descending from the maximal threshold $p \geq 1$, this amounts to computing the homology of a nested sequence of binary segmentations: $\tilde{S}_1 \subset ... \subset \tilde{S}_{p} \subset ... \subset \tilde{S}_{-\infty}$.
The barcode, therefore, is a dynamic characterisation of probabilistic segmentation topology, tracking its evolution against this threshold, $p$.
An exemplar sequence is demonstrated in \autoref{fig:PH_barcode}.
At each threshold, the number of vertically arranged bars indicates the topological features of $\tilde{S}_{p}$.
The presentation of each bar reveals the dimensionality of the feature described: solid bars indicate connected components; open bars indicate loops.
As such, the Betti numbers of $\tilde{S}_{p}$ are returned by counting the number of each type of bar vertically intersected.

A bar extends horizontally over the probability interval for which its associated feature is maintained.
Critical values of $p$ admit changes in the topological features of $\tilde{S}_{p}$ and are indicated by the endpoints of each bar.
Accordingly, the persistence of a topological feature $\Delta p$ is the horizontal breadth of its associated bar.
Persistent bars are considered robust to small perturbations, suggesting that they are true topological features of $\tilde{S}$.
Hence, in \autoref{fig:PH_barcode}, we arrange bars in order of descending persistence after grouping by topological dimension.
From the persistence barcode of the probabilistic segmentation $\tilde{S}$, we write the persistence of the $l^{th}$ most persistent feature of dimension $d$ as $\Delta p_{d,l}(\tilde{S})$.
Topological persistence is a differentiable quantity that enables gradient-based learning \cite{Bruel-Gabrielsson2019,Hofer2017}.

\section{Methods}
\label{sec:methods}

\subsection{Notation}

We address the generic multi-class segmentation task, seeking a meaningful division of the $N$D CMR image $\mathbf{X}: \mathbb{R}^{N} \rightarrow \mathbb{R}$ into meaningful anatomical segments.
In the tasks considered, image segments describe the chambers of the heart, their association, and their relationship to the myocardium (see \autoref{fig:tasks}).
We use the following shorthand: myocardium ($my$), left atrium ($la$), left ventricle ($lv$), right atrium ($ra$) and right ventricle ($rv$).
We denote the ground truth image segmentation by $\mathbf{Y}: \mathbb{R}^{N}\rightarrow \{ 0, 1 \} ^{K}$, being made up by $K$ mutually exclusive class label maps: $Y_{1}$, $Y_{2}$, ... $Y_{K}$, including $K-1$ foreground classes.
We consider $Y_{1}$ the background.

In each task we optimise the parameters, $\theta$, of a CNN to infer the probabilistic segmentation, $\mathbf{\tilde{Y}}: \mathbb{R}^{N}\rightarrow [ 0, 1 ] ^ K$,
a distribution over per-class segmentation maps: $\tilde{Y}_{1}$, $\tilde{Y}_{2}$, ... $\tilde{Y}_{K}$.
We write inference as $\mathbf{\tilde{Y}}=f(\mathbf{X};\mathbf{\theta})$.
In all cases, we achieve a discrete prediction as the segmentation which maximises pixel-wise probability: $\mathbf{\hat{Y}}: \mathbb{R}^{N}\rightarrow \{ 0, 1 \} ^ K$, with $K$ mutually exclusive classes: $\hat{Y}_{1}$, $\hat{Y}_{2}$, ... $\hat{Y}_{K}$.
Given the success of CNN-based solutions, we assume that, at least with respect to spatial overlap, $\mathbf{\hat{Y}}$ is a reasonable estimate of $\mathbf{Y}$.

Within our formalism, we consider the topology, not only of individual segmentation objects, but also their combination: by $Y_{i \cup j}$ and $\hat{Y}_{i \cup j}$ we refer to the Boolean union of classes $i$ and $j$; $\tilde{Y}_{i \cup j}$ is the pixel-wise probability of class $i$ or $j$.
We consider the union of a class with itself to be the segmentation of the single class: $Y_{i \cup j=i} = Y_{i}$, $\hat{Y}_{i \cup j=i} = \hat{Y}_{i}$ and $\tilde{Y}_{i \cup j=i} = \tilde{Y}_{i}$.

\subsection{Multi-class topological priors}
\label{subsec:mc_topo_priors}

The Betti numbers are topological invariants permitting the specification of priors for the description of foreground image segments.
Consider our 2D, short axis example (see \autoref{fig:tasks}):
\begin{alignat*}{12}
  &\mathbf{b}^{rv} &&= (1, 0) \hspace{4mm} &&\textrm{(1)}\stepcounter{equation} \hspace{16mm} &&\mathbf{b}^{rv \cup my} &&= (1, 1) \hspace{4mm} &&\textrm{(4)}\stepcounter{equation} \\
  &\mathbf{b}^{my} &&= (1, 1) \hspace{4mm} &&\textrm{(2)}\stepcounter{equation} \hspace{16mm} &&\mathbf{b}^{rv \cup lv} &&= (2, 0) \hspace{4mm} &&\textrm{(5)}\stepcounter{equation} \\
  &\mathbf{b}^{lv} &&= (1, 0) \hspace{4mm} &&\textrm{(3)}\stepcounter{equation} \hspace{16mm} &&\mathbf{b}^{my \cup lv} &&= (1, 0) \hspace{4mm} &&\textrm{(6)}\stepcounter{equation}
\end{alignat*}
Equations (1)-(3) specify that the $rv$, $my$ and $lv$ each comprise a single connected component, and that the $my$ contain a loop.
However, these equations only provide a topological specification in a segment-wise fashion: they fail to capture inter-class relationships.
For instance, they make no specification that the $my$ surround the $lv$ or that the $rv$ and $my$ be adjacent.
\todo{R2.3a}By the inclusion-exclusion principle, the topology of a 2D multi-class segmentation is \markup{theoretically} determined by all foreground objects and all object pairs \cite{Bazin2007a}.
Hence, a richer topological description also considers the combined foreground classes set out in Equations (4)-(6).

For convenience we collect Equations (1)-(6) into a task-specific Betti array $\mathbf{B}: \{1, 2, 3\} \times \{1, 2, 3\} \times \{0, 1\} \rightarrow \mathbb{R}$.
Each element $B^{ij}_{d}$ denotes the Betti number of dimension $d$ for the ground truth segmentation $Y_{i \cup j}$.
Vitally, even in the absence of the ground truth, $\mathbf{B}$ can be determined by prior knowledge of the anatomy to be segmented.

\todo{R2.2a R3.8}\markup{In 3D, theoretical requirement extends topological specification to also include all possible object triples.
For the segmentation of 3D whole heart anatomy from volumetric CMR (see \autoref{fig:tasks}(b)), this presents a rich prior description:
\begin{alignat*}{12}
&\mathbf{b}^{my}         &&= (1, 0, 0) \hspace{4mm} &&\textrm{(7)} \stepcounter{equation} \hspace{10mm}&&\mathbf{b}^{my \cup la \cup lv} &&= (1, 0, 0) \hspace{4mm} &&\textrm{(22)}\stepcounter{equation} \\
&\mathbf{b}^{la}         &&= (1, 0, 0) \hspace{4mm} &&\textrm{(8)} \stepcounter{equation} \hspace{10mm}&&\mathbf{b}^{my \cup la \cup ra} &&= (1, 0, 0) \hspace{4mm} &&\textrm{(23)}\stepcounter{equation} \\
&\mathbf{b}^{lv}         &&= (1, 0, 0) \hspace{4mm} &&\textrm{(9)} \stepcounter{equation} \hspace{10mm}&&\mathbf{b}^{my \cup la \cup rv} &&= (1, 0, 0) \hspace{4mm} &&\textrm{(24)}\stepcounter{equation} \\
&\mathbf{b}^{ra}         &&= (1, 0, 0) \hspace{4mm} &&\textrm{(10)}\stepcounter{equation} \hspace{10mm}&&\mathbf{b}^{my \cup lv \cup ra} &&= (1, 0, 0) \hspace{4mm} &&\textrm{(25)}\stepcounter{equation} \\
&\mathbf{b}^{rv}         &&= (1, 0, 0) \hspace{4mm} &&\textrm{(11)}\stepcounter{equation} \hspace{10mm}&&\mathbf{b}^{my \cup lv \cup rv} &&= (1, 0, 0) \hspace{4mm} &&\textrm{(26)}\stepcounter{equation} \\
&                        &&                         &&                                                 &&\mathbf{b}^{my \cup ra \cup rv} &&= (1, 0, 0) \hspace{4mm} &&\textrm{(27)}\stepcounter{equation} \\
&\mathbf{b}^{my \cup la} &&= (1, 0, 0) \hspace{4mm} &&\textrm{(12)}\stepcounter{equation} \hspace{10mm}&&\mathbf{b}^{la \cup lv \cup ra} &&= (2, 0, 0) \hspace{4mm} &&\textrm{(28)}\stepcounter{equation} \\
&\mathbf{b}^{my \cup lv} &&= (1, 0, 0) \hspace{4mm} &&\textrm{(13)}\stepcounter{equation} \hspace{10mm}&&\mathbf{b}^{la \cup lv \cup rv} &&= (2, 0, 0) \hspace{4mm} &&\textrm{(29)}\stepcounter{equation} \\
&\mathbf{b}^{my \cup ra} &&= (1, 0, 0) \hspace{4mm} &&\textrm{(14)}\stepcounter{equation} \hspace{10mm}&&\mathbf{b}^{la \cup ra \cup rv} &&= (2, 0, 0) \hspace{4mm} &&\textrm{(30)}\stepcounter{equation} \\
&\mathbf{b}^{my \cup rv} &&= (1, 0, 0) \hspace{4mm} &&\textrm{(15)}\stepcounter{equation} \hspace{10mm}&&\mathbf{b}^{lv \cup ra \cup rv} &&= (2, 0, 0) \hspace{4mm} &&\textrm{(31)}\stepcounter{equation} \\
&\mathbf{b}^{la \cup lv} &&= (1, 0, 0) \hspace{4mm} &&\textrm{(16)}\stepcounter{equation} \\
&\mathbf{b}^{la \cup ra} &&= (2, 0, 0) \hspace{4mm} &&\textrm{(17)}\stepcounter{equation} \\
&\mathbf{b}^{la \cup rv} &&= (2, 0, 0) \hspace{4mm} &&\textrm{(18)}\stepcounter{equation} \\
&\mathbf{b}^{lv \cup ra} &&= (2, 0, 0) \hspace{4mm} &&\textrm{(19)}\stepcounter{equation} \\
&\mathbf{b}^{lv \cup rv} &&= (2, 0, 0) \hspace{4mm} &&\textrm{(20)}\stepcounter{equation} \\
&\mathbf{b}^{ra \cup rv} &&= (1, 0, 0) \hspace{4mm} &&\textrm{(21)}\stepcounter{equation} \\
\end{alignat*}}

\subsection{Topological loss function}

\begin{figure}[t]
\centering
\includegraphics[width=0.9\linewidth]{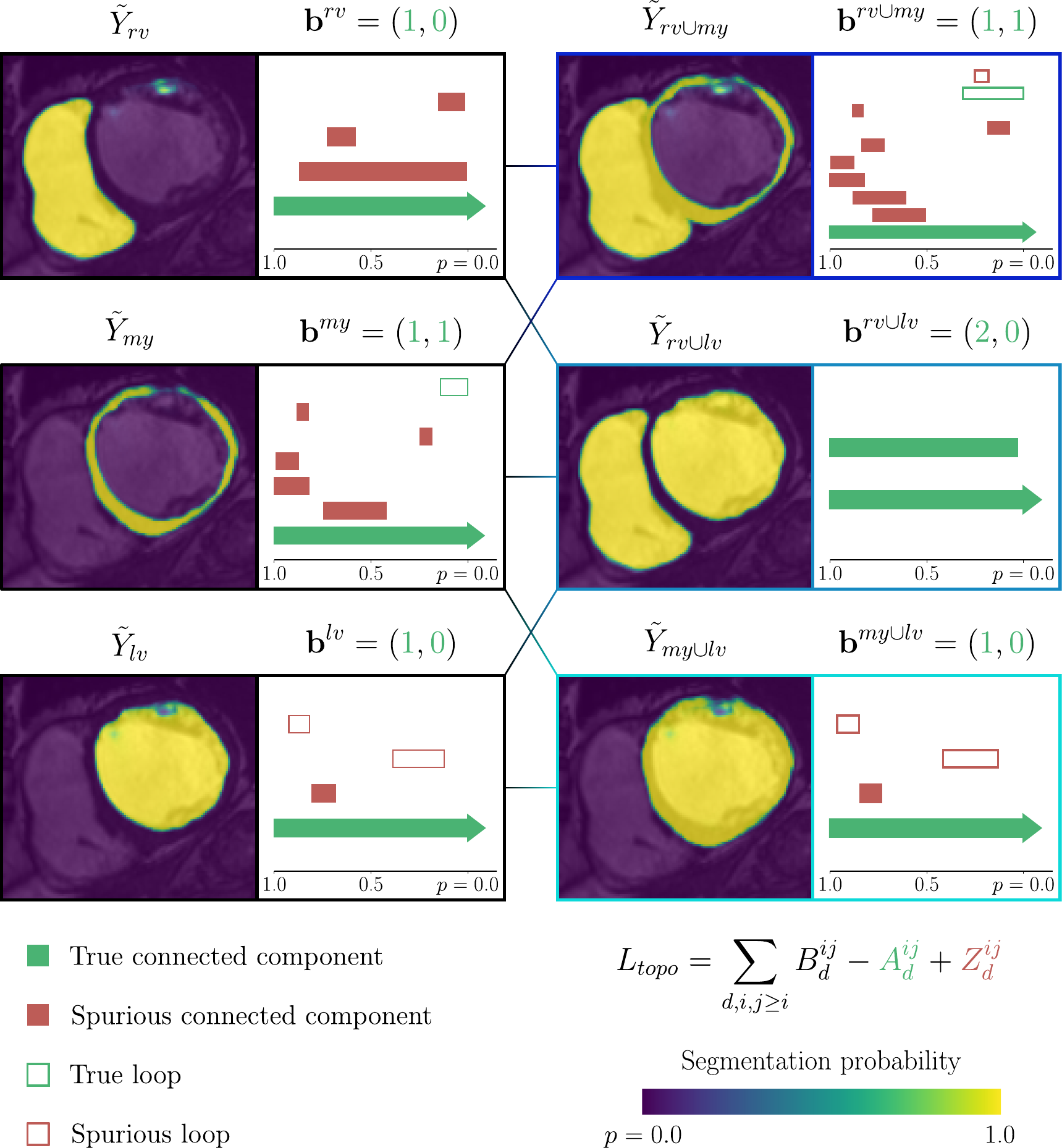}
\caption{Construction of the loss $L_{topo}$.
Each probabilistic segmentation ($\tilde{Y}_{i}$ or $\tilde{Y}_{i \cup j}$), is accompanied by its associated persistence barcode (only bars with $\Delta p_{d,l} \ge 0.05$ are shown).}
\label{fig:L_topo}
\end{figure}

\todo{R1.1a}\markup{Though alternatives have been presented (see \autoref{subsec:form_comp}), the underlying basis to our loss functions exploits the differentiable properties of the persistence barcode to expose the differences between $\mathbf{\tilde{Y}}$ and our prior specification $\mathbf{B}$.}
However, in contrast with previous works \cite{Clough2020,Shin2020}, which considered binary segmentation problems, our loss is informed by the topological features of several probabilistic segmentations.
Critically, segmentation selection and construction defines the topological features to which our formulation is sensitive.
Given the theory of \autoref{subsec:mc_topo_priors}, we consider the persistence barcode for all foreground class labels and class label pairs (see Figure \ref{fig:L_topo}):
\begin{align}
\label{eq:L_topo}
L_{topo}   &= \; \sum_{d, i, j \geq i} B_{d}^{ij} - A^{ij}_{d} + Z^{ij}_{d}
\end{align}
\vspace{-3mm}
\begin{equation*}
    A^{ij}_{d} = \sum_{l=1}^{B_{d}^{ij}} \Delta p_{d,l}(\tilde{Y}_{i \cup j}) \quad \quad \quad \quad
    Z^{ij}_{d} = \!\!\!\!\!\sum_{l=B_{d}^{ij} + 1}^{\infty} \!\!\!\!\! \Delta p_{d,l}(\tilde{Y}_{i \cup j})
\end{equation*}
Recalling that $\Delta p_{d,l}$ is the persistence of the $l^{th}$ most persistent topological feature of dimension $d$, $A^{ij}_{d}$ evaluates the total persistence of the $B_{d}^{ij}$ longest $d$-dimensional bars for the probabilistic union of classes $i$ and $j$, $\tilde{Y}_{i \cup j}$.
Assuming that the inferred segmentation closely approximates the ground truth, $A^{ij}_{d}$ measures the presence of anatomically meaningful topological features: the extent to which the topological features anticipated by $B^{ij}_{d}$ are present within $\mathbf{\tilde{Y}}$.
$Z^{ij}_{d}$ evaluates the persistence of features that are superfluous to $B_{d}^{ij}$: it is sensitive to spurious connected components or holes.
Summing over all topological dimensions $d$, and considering all classes $i,j=i$ and class pairs $i,j>i$, optimising $L_{topo}$ maximises the persistence of topological features which match the prior specification, and minimises those which do not.

\subsection{CNN-based post-processing framework}

\markup{Though consistent with fully or weakly supervised training,} we\todo{R2.7} demonstrate our topological loss function in a CNN post-processing framework \cite{Byrne2021,Clough2020}, seeking an improvement in segmentation topology by fine tuning a pre-trained CNN, $f(\mathbf{X}; \theta)$.
This achieves a new set of network parameters $\theta_{n}$, optimised to correct the topology of the specific test case $\mathbf{X}_{n}$.

However, since topology is a global property, many segmentations potentially minimise $L_{topo}$.
Hence, where $V_{n}$ is the number of pixels in $\mathbf{X}_{n}$, $L_{mse}$ limits test time adaptation to the minimal set of modifications necessary to align the segmentation and the topological prior, $\mathbf{B}$.
The hyperparameter $\lambda$ controls the influence of the associated similarity constraint:
\begin{align}
L_{mse} &=  \frac{1}{V_{n}} |f(\mathbf{X}_{n}; \theta) - f(\mathbf{X}_{n}; \theta_{n})|^{2} \\
\label{eq:ltp}
L_{TP}  &= L_{topo}(f(\mathbf{X}_{n}; \theta_{n}), \mathbf{B}) + \lambda L_{mse}
\end{align}

\subsection{Computation of the PH barcode}
\label{subsec:implementation}

Computation of the persistence barcode is an intensive procedure, introducing significant inefficiencies within CNN training \cite{Shin2020}.
This is particularly problematic in 3D and within our loss formulation, demanding the extraction of multiple barcodes per gradient update.
In this work, we have implemented a thin wrapper around the open source ``CubicalRipser" library \cite{Kaji2020}, integrating its functionality with the automatic differentiation engine of PyTorch.
As discussed in \autoref{sec:discussion}, this brings about dramatic performance improvements when compared with previous works \cite{Clough2020,Byrne2021} reliant on the ``TopologyLayer" library \cite{Bruel-Gabrielsson2019}.
\todo{R2.4}\markup{Moreover, we use multiprocessing from the Python Standard Library to make multiple calls to CubicalRipser to compute barcodes in parallel.}

\section{Experiments}

\subsection{Experimental methods and baselines}
\label{subsec:baselines}

We compare performance with the following baselines: (1) $U$-$Net$, The discrete segmentation maximising pixel-wise probability, conventionally inferred by U-Net; and (2)
$CCA$, Connected component analysis: the discrete segmentation composed by the per-class largest components of $U$-$Net$.

We test the performance of two flavours of topological post-processing.
Both deploy a variant of \autoref{eq:ltp}:
(1) $TP_{i,j=i}$, post-processing according to per-class topological priors specified for individual foreground labels only; and (2) $TP_{i,j \ge i}$, post-processing according to multi-class topological priors, specified for all individual and paired foreground labels.

\subsection{Metrics and statistical analysis}
\label{subsec:metrics}

\todo{R3.2a}\markup{Expressed as a percentage, we use the per class DSC as an objective measure of spatial overlap performance.
We summarise over classes using the generalised DSC, $gDSC$ \cite{Zhuang2019}, and note the change in spatial overlap induced by post-processing as $\Delta gDSC$.
\todo{R3.4a}We use the per class Hausdorff distance ($HDD$) to assess the accuracy of surface localisation.}

We characterise topological accuracy using two metrics.
The Betti error ($BE$) measures the total deviation between \markup{the topology of the discrete predicted segmentation ($\hat{\mathbf{b}}$) and the ground truth specification ($\mathbf{b}$).\todo{R2.5}
In 2D, $BE$ considers all individual and paired foreground classes:
\begin{equation} \label{eq:be2d}
BE = \sum_{i, j \ge i} \| \hat{\mathbf{b}}^{i \cup j} - \mathbf{b}^{i \cup j} \|_{1}
\end{equation}}
In 3D, $BE$ also considers class triples, $i\cup j \cup k$.
To understand whether improvements in segmentation topology translate into anatomically meaningful segmentations we also make a binary assessment of topological success ($TS$):
\begin{equation} \label{eq:sr}
TS = \begin{cases}
			1, & \text{if $BE = 0$}\\
            0, & \text{otherwise}
		 \end{cases}
\end{equation}
\noindent

Spatial overlap results are described by their median and interquartile range.
To account for large differences in performance, Betti error is described by the $i^{th}$ percentile ($P_i$) and percentile ranges indicated.
Results for $gDSC$ and $BE$ are compared using Wilcoxon signed rank test with Bonferroni correction.
\todo{R3.6}\markup{$TS$ is summarised by the percentage of test cases for which a positive result was achieved} ($\rho$) and its associated standard deviation ($\sigma_{\rho}$), and compared using exact binomial test after Bonferroni correction.

\subsection{Experiment 1: 2D short axis segmentation}
\label{subsec:short_axis}

\begin{table*}[t]
\centering
\caption{Aspects of spatial coherence on the test set, held out from the ACDC training data: segmentation of 2D short axis CMR.}\label{tab:topo_2D}
\markup{\begin{tabularx}{\textwidth}{*{1}{L}|*{3}{L}|*{2}{L}}
\toprule
&
    \multicolumn{3}{c|}{\textbf{2D per class Hausdorff distance (mm)}} &
    \multicolumn{2}{c}{\textbf{Topological performance}} \\
&
    $HDD_{rv}$ &
    $HDD_{my}$ &
    $HDD_{lv}$
    & $BE$
    & $TS$ (\%) \\
\toprule
$U$-$Net$ &
    4.75$_{\scaleto{\text{(3.54,7.91)}}{\iqr}}$ &
    3.54$_{\scaleto{\text{(2.80,4.51)}}{\iqr}}$ &
    2.80$_{\scaleto{\text{(2.50,3.75)}}{\iqr}}$ &
    5.51$_{\scaleto{\text{(4.02,8.00)}}{\iqr}}$ &
    85.3$_{\scaleto{\text{(35.4)}}{\iqr}}$ \\
\midrule
$+ CCA$ &
    4.51$_{\scaleto{\text{(3.54,7.60)}}{\iqr}}$ &
    3.54$_{\scaleto{\text{(2.80,3.95)}}{\iqr}}$ &
    2.80$_{\scaleto{\text{(2.50,3.75)}}{\iqr}}$ &
    3.02$_{\scaleto{\text{(2.00,4.00)}}{\iqr}}$ &
    92.7$_{\scaleto{\text{(26.0)}}{\iqr}}$ \\
\midrule
$+ {TP}_{i, j = i}$ &
    4.51$_{\scaleto{\text{(2.80,7.60)}}{\iqr}}$ &
    3.54$_{\scaleto{\text{(2.57,3.95)}}{\iqr}}$ &
    2.50$_{\scaleto{\text{(2.50,3.70)}}{\iqr}}$ &
    1.00$_{\scaleto{\text{(1.00,2.00)}}{\iqr}}$ &
    97.3$_{\scaleto{\text{(16.2)}}{\iqr}}$ \\
$+ {TP}_{i, j \geq i}$ &
    4.51$_{\scaleto{\text{(2.80,7.60)}}{\iqr}}$ &
    3.54$_{\scaleto{\text{(2.50,3.95)}}{\iqr}}$ &
    2.50$_{\scaleto{\text{(2.50,3.70)}}{\iqr}}$ &
    0.51$_{\scaleto{\text{(0.00,1.00)}}{\iqr}}$ &
    98.7$_{\scaleto{\text{(11.3)}}{\iqr}}$ \\
\bottomrule
\multicolumn{6}{l}{\scriptsize Reporting:
    $HDD_{\{rv,my,lv\}}$ $P_{50_{\scaleto{(P_{25}, P_{75})} {\iqr}}}$;
    $BE$ $P_{99_{\scaleto{(P_{98}, P_{100})}{\iqr}}}$;
    $TS$ $\rho_{\scaleto{(\sigma_{\rho})}{\iqr}}$.}
\end{tabularx}}\todo{\normalfont R3.4b}
\end{table*}

\subsubsection{Experimental setting}

In this experiment we demonstrate our approach on multi-class, 2D short axis segmentation, using a subset of the publicly available ACDC training dataset \cite{Bernard2018}.
\todo{R3.5a}\markup{To avoid topological changes observed at apex and base, and permit the specification of a uniform prior}, we extract three mid ventricular slices from each short axis stack, including diastolic and systolic frames from all 100 patients\markup{: 600 images in total.}
All image-label pairs were resampled to an isotropic spacing of 1.25~mm and normalised to have zero mean and unit variance \cite{Isensee2018}.
Subjects were randomly divided between training, validation and test sets in the ratio 2:1:1, stratified by ACDC diagnostic classification.
We trained a 2D U-Net \cite{Ronneberger2015} using CE loss and the combined training and validation set of 450 images, for 16,000 iterations.
We employed the Adam optimiser with a learning rate of $10^{-3}$.
Each minibatch contained ten patches of size 352~by~352, randomly cropped from ten different patients.
Data augmentation applied random rotation.

Topological post-processing was performed on the inferred multi-class segmentations of the held-out test set.
This sought to minimise $L_{TP}$ for the \markup{prior expressed by equations (1-6)}, with $\lambda = 1000$.
Test time adaptation used the Adam optimiser with a learning rate of $10^{-5}$ for 100 iterations.
Hyperparameters for both supervised training and topological post-processing were optimised using the validation set.

\subsubsection{Results}

\begin{table*}[b]
\centering
\caption{Spatial overlap performance on the test set, held out from the ACDC training data: segmentation of 2D short axis CMR.
}\label{tab:dsc_2D}
\markup{\begin{tabularx}{\textwidth}{*{1}{L}|*{3}{L}|*{2}{L}}
\toprule
&
    \multicolumn{3}{c|}{\textbf{Per class performance} (\%)} &
    \multicolumn{2}{c}{\textbf{Generalised performance} (\%)} \\
&
    $DSC_{rv}$ &
    $DSC_{my}$ &
    $DSC_{lv}$ &
    $gDSC$ &
    $\Delta gDSC$ \\ 
\toprule
$U$-$Net$ &
    92.8$_{\scaleto{\text{(88.2,95.1)}}{\iqr}}$ &
    90.1$_{\scaleto{\text{(87.2,92.7)}}{\iqr}}$ &
    96.1$_{\scaleto{\text{(94.7,97.1)}}{\iqr}}$ &
    93.2$_{\scaleto{\text{(91.7,94.3)}}{\iqr}}$ &
    --- \\
\midrule
$+CCA$ &
    92.8$_{\scaleto{\text{(88.2,95.1)}}{\iqr}}$ &
    90.2$_{\scaleto{\text{(87.3,92.7)}}{\iqr}}$ &
    96.1$_{\scaleto{\text{(94.7,97.1)}}{\iqr}}$ &
    93.2$_{\scaleto{\text{(91.7,94.3)}}{\iqr}}$ &
    $+$0.0$_{\scaleto{\text{(0.0,0.0)}}{\iqr}}$ \\
\midrule
$+{TP}_{i, j = i}$ &
    92.8$_{\scaleto{\text{(88.2,95.1)}}{\iqr}}$ &
    90.2$_{\scaleto{\text{(87.6,92.5)}}{\iqr}}$ &
    96.1$_{\scaleto{\text{(94.8,97.1)}}{\iqr}}$ &
    93.4$_{\scaleto{\text{(91.6,94.3)}}{\iqr}}$ &
    $+$0.0$_{\scaleto{\text{(0.0,0.1)}}{\iqr}}$ \\
$+{TP}_{i, j \geq i}$ &
    92.9$_{\scaleto{\text{(88.3,95.0)}}{\iqr}}$ &
    90.1$_{\scaleto{\text{(87.6,92.5)}}{\iqr}}$ &
    96.2$_{\scaleto{\text{(94.7,97.1)}}{\iqr}}$ &
    93.3$_{\scaleto{\text{(91.7,94.3)}}{\iqr}}$ &
    $+$0.0$_{\scaleto{\text{(0.0,0.1)}}{\iqr}}$ \\
\bottomrule
\multicolumn{6}{l}{\scriptsize Reporting:
    $DSC_{\{rv,my,lv\}}$ $P_{50_{\scaleto{(P_{25}, P_{75})} {\iqr}}}$;
    $gDSC$ $P_{50_{\scaleto{(P_{25}, P_{75})} {\iqr}}}$;
    $\Delta gDSC$ $P_{50_{\scaleto{(P_{25}, P_{75})} {\iqr}}}$.}
\end{tabularx}}\todo{\normalfont R2.8, R3.2b}
\end{table*}

\begin{figure}[t]
\minipage{0.29\linewidth}
  \centering
  \includegraphics[width=\linewidth]{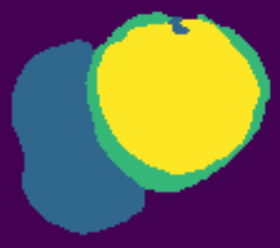}
  (a) $UNet$
\endminipage\hfill
\minipage{0.29\linewidth}
  \centering
  \includegraphics[width=\linewidth]{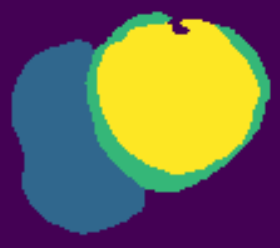}
  (b) $CCA$
\endminipage\hfill
\minipage{0.29\linewidth}
  \centering
  \includegraphics[width=\linewidth]{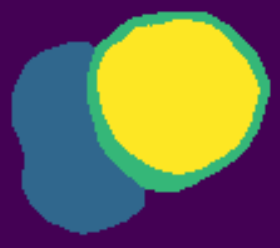}
  (c) $TP_{i, j \geq i}$
\endminipage
\caption{Where CCA is limited to the discrete removal of superfluous regions,
topological post-processing rectifies errors by expressive modification.
(c) suppression of the anomalous $rv$ component is accompanied by consolidation of the $lv$ cavity and completion of the myocardial torus.
}
\label{fig:short_axis_expressive}
\end{figure}
\begin{figure}[t]
\minipage{0.29\linewidth}
  \centering
  \includegraphics[width=\linewidth]{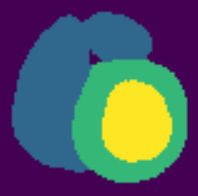}
  (a) $UNet$
\endminipage\hfill
\minipage{0.29\linewidth}
  \centering
  \includegraphics[width=\linewidth]{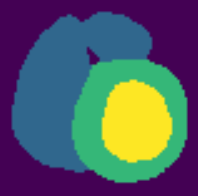}
  (b) $TP_{i, j = i}$
\endminipage\hfill
\minipage{0.29\linewidth}
  \centering
  \includegraphics[width=\linewidth]{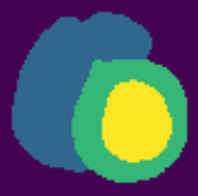}
  (c) $TP_{i, j \geq i}$
\endminipage\hfill
\caption{Specification of topological priors for all pairs of foreground classes captures the interaction between classes.
}
\label{fig:short_axis_incremental}
\end{figure}

Row one of \autoref{tab:topo_2D} demonstrates that U-Net segmentations exhibit topological errors in approximately 15\% of cases.
\markup{Also considering \autoref{tab:dsc_2D}, note that} these failings present despite strong \markup{$HDD$ and $DSC$}\todo{R3.4d} performance that is consistent with the state of the art.
This suggests that measures of \markup{surface localisation} and spatial overlap do not reliably predict topological performance.

It is apparent from row two of \autoref{tab:topo_2D} that approximately two fifths of individual Betti errors incurred by U-Net are resolved by rudimentary CCA ($p = 0.012$).
This translates into a significant increase in $TS$ ($p = 0.006$), accounting for approximately one half of the cases predicted with incorrect topology by U-Net.
Sensitive only to 0D topological features, CCA cannot rectify errors related to the presence of loops.
Moreover, as most frequently employed, CCA takes a discrete approach to segmentation post-processing considering connected components of the thresholded
segmentation.
In contrast, our topological loss functions modify probabilistic segmentation topology via CNN optimisation and are conditioned on the test image in question.
This admits topological improvement via expressive modification (see \autoref{fig:short_axis_expressive}).

The superior topological performance of our loss functions is quantified in \autoref{tab:topo_2D}.
Row three reflects the naive extension of previous work \cite{Clough2019} to the multi-class setting, specifying a single prior per class: $TP_{i,j = i}$.
This setup admits improved topological performance compared with CCA, resolving all topological errors in over 97\% of cases.
This reflects substantial but insignificant improvements in the average $BE$ ($p = 0.069$), and $TS$ ($p = 0.393$), compared against CCA.

Row four outlines the performance of the full realisation of our approach, specifying topological priors per class and per pairwise combination of classes: $TP_{i,j \ge i}$.
This provides topologically accurate results in almost 99\% of test images, failing in just two, and returns the lowest $BE$.
Compared with the naive specification of per-class topological priors only, pairwise combinations expose a richer description of multi-class segmentation topology (see \autoref{fig:short_axis_incremental}).
Compared with CCA,  $TP_{i,j \ge i}$ makes significant improvements in $BE$ ($p = 0.026$) and $TS$ ($p = 0.023$).

\begin{figure}[t]
\centering
\includegraphics[width=0.8\linewidth]{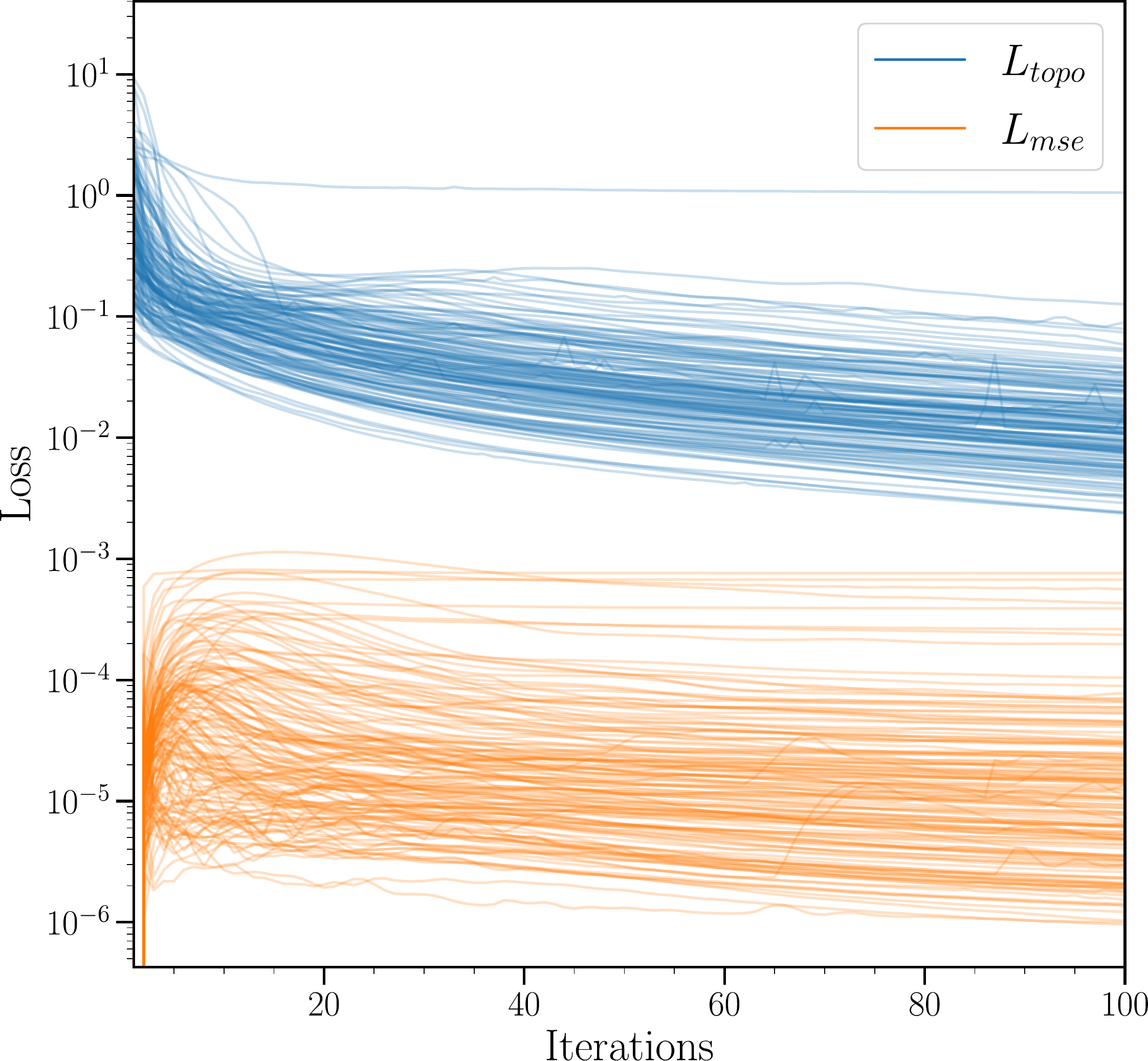}
\caption{Optimisation of $L_{TP}$ in the task of 2D short axis segmentation for the entire ACDC test set.
}
\label{fig:short_axis_loss}
\end{figure}

\autoref{fig:short_axis_loss} illustrates the evolution of the different terms of \autoref{eq:ltp}, demonstrating that the smooth improvement in $L_{topo}$ is balanced by a small increase in the similarity constraint $L_{mse}$.
The latter ensures that post-processed segmentations cannot deviate dramatically from the initial U-Net prediction and maintains spatial overlap performance.
\markup{This is reflected in \autoref{tab:dsc_2D}\todo{R3.2d R3.4d}:
on average, all post-processing methods marginally improved per class $DSC$, $HDD$ and $gDSC$ over U-Net.}
However, any gain made was not substantial when compared with the DSC between manual segmentations of different observers: estimated to be around 0.9 \cite{Bai2018}.

\subsection{Experiment 2: 3D whole heart segmentation}

In this experiment, we apply our methods to the task of segmenting multi-class, whole heart anatomy from isotropic, high spatial resolution image data.
We consider a semantic subset of the MM-WHS Challenge task, seeking a segmentation of the CMR volume into $la$, $ra$, $lv$, $rv$ and $my$ \cite{Zhuang2019}.
In the context of topological optimisation, this task is made particularly challenging by the semantics of the publicly available ground truth training data.
Primarily concerned with measures of spatial overlap and surface error, these segmentations need not and do not convey clinically meaningful topology\todo{3.7b}\markup{, failing to describe the haemodynamic continuity and isolation of the heart's chambers by pixel adjacency (see \autoref{fig:tasks}(b))}.
Hence, this application of our method seeks to not only refine, but also to \emph{impose} meaningful segmentation topology.

\todo{R2.2b R2.3b}\markup{Theoretically, a 3D topological prior should reflect all foreground classes, their pair-wise combination, and their collection in all possible groups of three.
However, for pragmatic reasons we overlook the topology of label triples.
Firstly, in 3D the number of barcodes computed at each topological post-processing iteration strongly influences execution time.
For five foreground classes, including label triplets almost doubles the number of barcodes from 15 to 25 (exceeding our parallel computing resources of 16 CPU cores; note that provided sufficient processors our parallelisation extends immediately to a theoretically founded prior for this and more complex tasks).
Secondly, the semantics of our task do not anticipate anatomically meaningful features involving the combination of three classes.
Accordingly, our specification by equations (7-21) reflects: (i) continuity of the atrioventricular junctions; (ii) isolation of the left and right heart; and (iii)
adjacency of the $rv$ blood pool and epicardium.
For evaluation, both $BE$ and $TS$ assess segmentation topology against equations (7-31).}

\label{subsec:whole_heart}
\subsubsection{Experimental setting}

We employed all twenty publicly available training cases and performance was assessed against the encrypted test set of forty examples.
Prior to experimentation, all data were resampled to an isotropic resolution of 1.00 mm and normalised to have zero mean and unit variance.
We elected to crop all volumes tightly around the foreground region of interest.
This allows us to maintain high spatial resolution (and sensitivity to topological features associated with thin tissue interfaces) without extreme increases in computational demand.

For our baseline we trained a 3D U-Net \cite{Cicek2016} using CE loss for 40000 iterations.
We employed the stochastic gradient descent optimiser, using a learning rate of 0.01 and momentum of 0.99.
Each minibatch contained two 3D image patches of size 192~by~160~by~160.
We made use of intensive data augmentation including rotation about all three spatial axes, scaling and non-rigid deformation.
Topological fine tuning was mediated by the Adam optimiser using a learning rate of $10^{-5}$ for 100 iterations and with $\lambda = 1$.
All hyperparameters were established by five fold cross-validation over the training set.

\subsubsection{Results}

\begin{table*}[t]
\centering
\caption{Performance on the encrypted MM-WHS test set: multi-class segmentation of high resolution 3D CMR.}
\label{tab:topo_3D}
\markup{\begin{tabularx}{\textwidth}{ *{1}{L} | *{5}{L} | *{2}{L} }
\toprule
&
    \multicolumn{5}{c|}{\textbf{3D per class Hausdorff distance (mm)}} &
    \multicolumn{2}{c}{\textbf{Topological performance}} \\
&
    $HDD_{la}$ &
    $HDD_{ra}$ &
    $HDD_{lv}$ &
    $HDD_{rv}$ &
    $HDD_{my}$ &
    $BE$ &
    $TS$ \\
\toprule
$U$-$Net$ &
    70.3$_{\scaleto{\text{(54.6,76.4)}}{\iqr}}$ &
    66.9$_{\scaleto{\text{(48.0,75.1)}}{\iqr}}$ &
    51.7$_{\scaleto{\text{(23.4,57.3)}}{\iqr}}$ &
    54.8$_{\scaleto{\text{(29.2,71.7)}}{\iqr}}$ &
    32.8$_{\scaleto{\text{(15.7,46.6)}}{\iqr}}$ &
    353$_{\scaleto{\text{(277,439)}}{\iqr}}$ &
    0.0$_{\scaleto{\text{(0.0)}}{\iqr}}$ \\
\midrule
$+CCA$ &
    \phantom{1}9.5$_{\scaleto{\text{(7.1,10.9)}}{\iqr}}$ &
    13.9$_{\scaleto{\text{(13.2,17.0)}}{\iqr}}$ &
    \phantom{1}9.9$_{\scaleto{\text{(8.2,12.2)}}{\iqr}}$ &
    13.7$_{\scaleto{\text{(11.0,16.4)}}{\iqr}}$ &
    14.2$_{\scaleto{\text{(12.3,15.8)}}{\iqr}}$ &
    \phantom{1}50$_{\scaleto{\text{(26,72)}}{\iqr}}$ &
    0.0$_{\scaleto{\text{(0.0)}}{\iqr}}$ \\
\midrule
$+{TP}_{i, j = i}$ &
    14.0$_{\scaleto{\text{(10.3,27.7)}}{\iqr}}$ &
    17.9$_{\scaleto{\text{(14.4,38.6)}}{\iqr}}$ &
    12.5$_{\scaleto{\text{(9.5,15.7)}}{\iqr}}$ &
    16.3$_{\scaleto{\text{(12.7,32.8)}}{\iqr}}$ &
    21.4$_{\scaleto{\text{(16.8,28.2)}}{\iqr}}$ &
    \phantom{1}11$_{\scaleto{\text{(7,18)}}{\iqr}}$ &
    0.0$_{\scaleto{\text{(0.0)}}{\iqr}}$ \\
$+{TP}_{i, j \geq i}$ &
    13.9$_{\scaleto{\text{(9.7,46.7)}}{\iqr}}$ &
    23.9$_{\scaleto{\text{(15.2,43.5)}}{\iqr}}$ &
    11.9$_{\scaleto{\text{(9.9,16.0)}}{\iqr}}$ &
    15.6$_{\scaleto{\text{(11.4,34.1)}}{\iqr}}$ &
    22.3$_{\scaleto{\text{(15.4,27.7)}}{\iqr}}$ &
    \phantom{11}5$_{\scaleto{\text{(2,8)}}{\iqr}}$ &
    5.0$_{\scaleto{\text{(21.8)}}{\iqr}}$ \\
\bottomrule
\multicolumn{8}{l}{\scriptsize Reporting:
    $HDD_{\{la,ra,lv,rv,my\}}$ $P_{50_{\scaleto{(P_{25}, P_{75})} {\iqr}}}$;
    $BE$ $P_{50_{\scaleto{(P_{25}, P_{75})}{\iqr}}}$;
    $TS$ $\rho_{\scaleto{(\sigma_{\rho})}{\iqr}}$.}
\end{tabularx}}\todo{\normalfont R3.4c}
\end{table*}

\begin{figure}[t]
\centering
\includegraphics[width=0.9\linewidth]{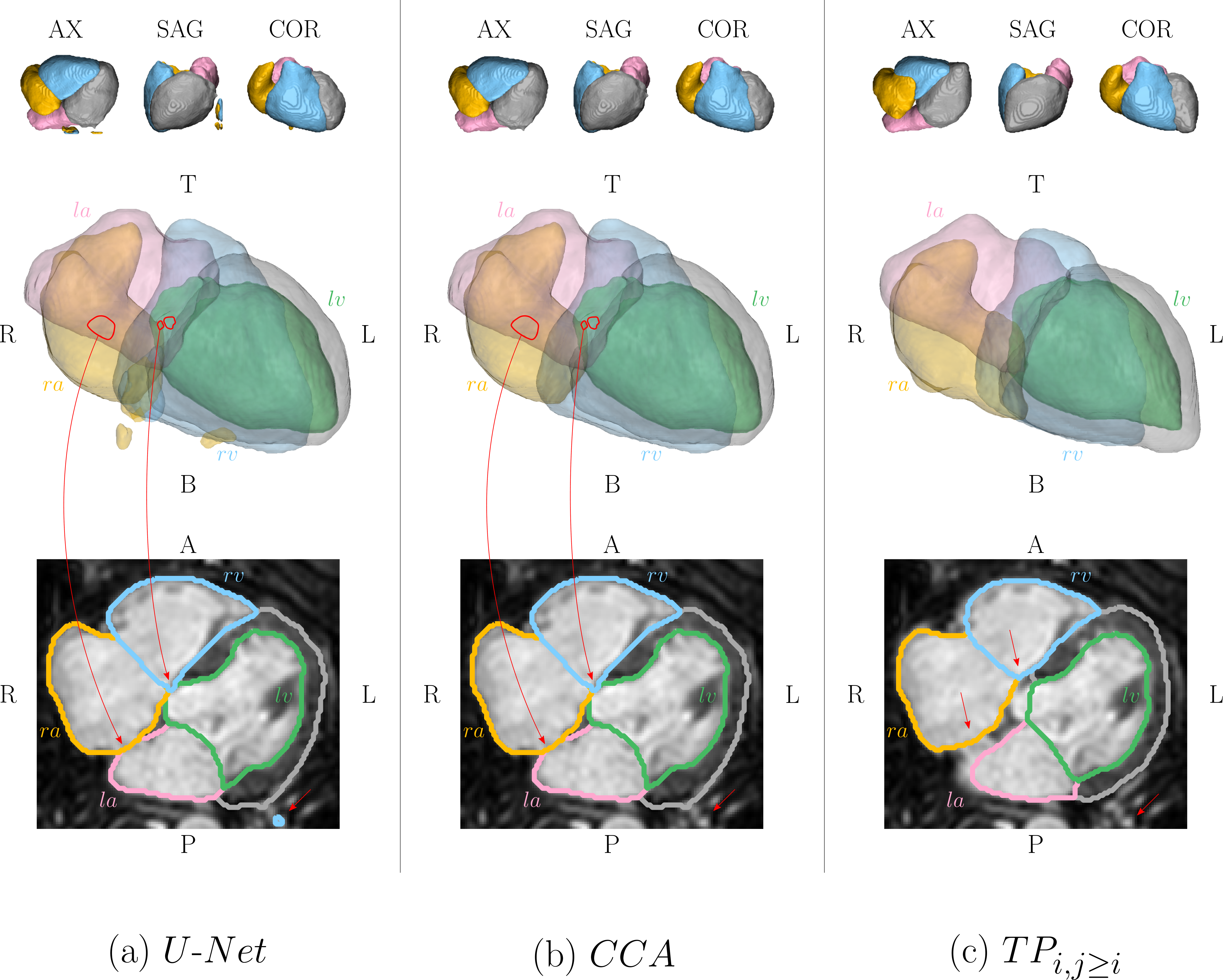}
\caption{In the setting of high $gDSC$ topological post-processing corrects segmentation topology by minimal adaptation.
(a) U-Net segmentation presents a number of topological errors.
(b) CCA successfully removes superfluous connected components, but the anomalous association of the left and right heart remains.
(c) $TP_{i, j \ge i}$ successfully corrects all errors.
Topological features of interest are indicated in red.}
\label{fig:wh_good}
\end{figure}
\begin{figure}[t]
\centering
\includegraphics[width=0.9\linewidth]{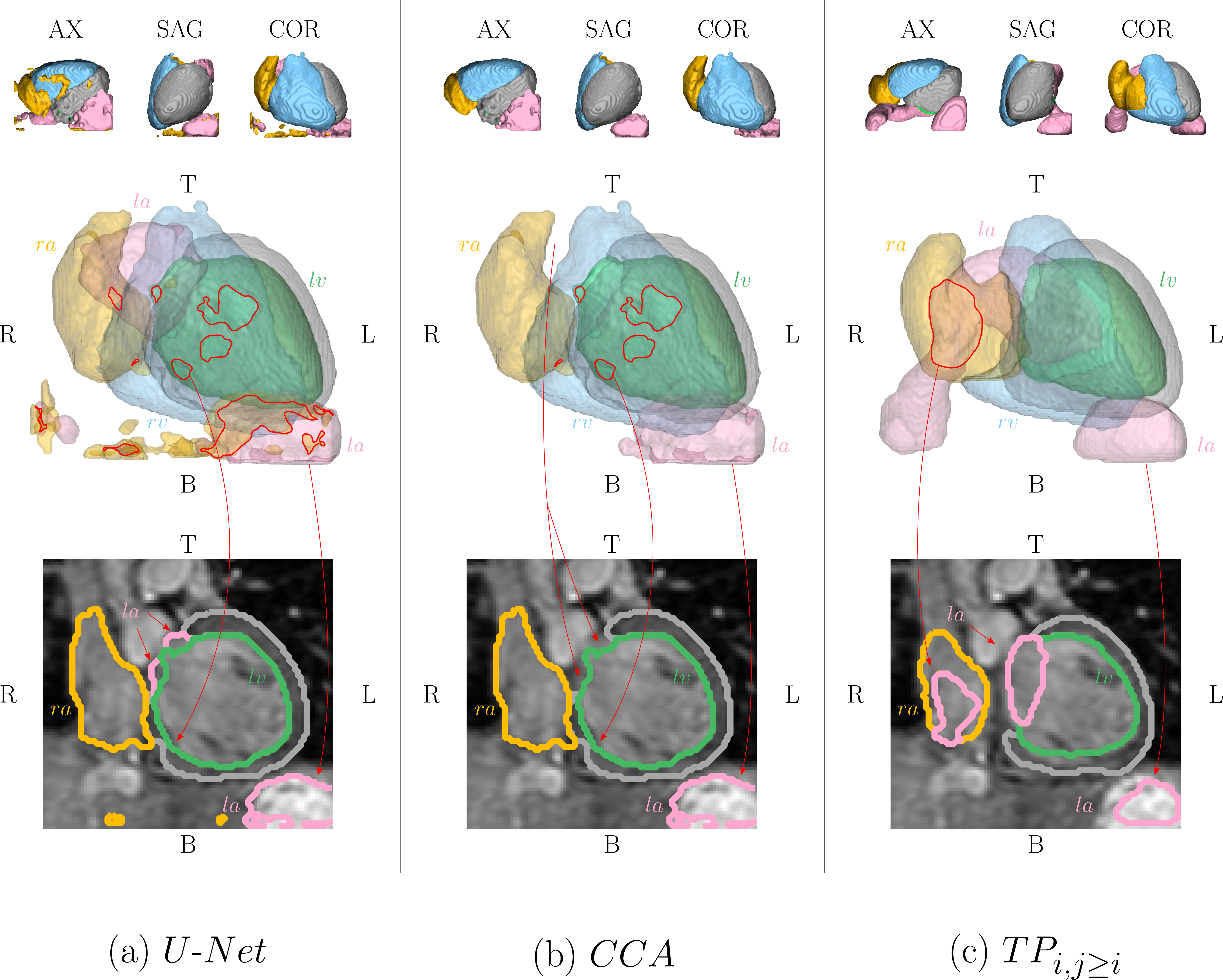}
\caption{The prediction of spurious connected components with high probability can confound topological post-processing.
(a) $U$-$Net$ segmentation includes significant spatial overlap and topological errors.
(b) Naive $CCA$ misidentifies the $la$.
(c) $TP_{i, j \ge i}$ is confounded by topologically credible $la$ candidates, maintaining an anomalous association with the $ra$.
Topological features of interest are indicated in red.}
\label{fig:wh_cc}
\end{figure}

\markup{Here, consideration of \autoref{tab:topo_3D} and} topological performance requires careful consideration.
Since the training data do not reflect clinically meaningful topology, it might be argued that evaluation against our prior has limited interpretability.
To inform this comparison, we examined the effective performance of the ground truth training labels.
Having a median $BE$ of $377$, these examples do not include a single topologically accurate segmentation.
\todo{R3.7c}\markup{Demonstrating that such features are not incidentally or easily achieved through conventional manual segmentation, they reinforce the need for topologically sensitive methods.}

It is clear from these results that the topology of the ground truth training data not only diverges from our clinically relevant specification, but is meaningless.
In this context, topological post-processing attempts to impose unseen topology at test time.
However, it is not alone in this regard: the same can be said for CCA.
Though widely employed as a practical approach to reduce false positives and improve spatial overlap, it must be acknowledged that here, CCA applies a limited, 0D topological prior, that conflicts with the features of training data.
Despite this, \autoref{tab:topo_3D} endorses the former rationalisation: coupled with a significant improvement in the average $gDSC$ ($p < 10^{-4}$), CCA modifies segmentation topology, reducing the $BE$ by just over 85\% ($p < 10^{-7}$).

However, CCA remains insensitive to both multi-class and high dimensional topology.
Sensitivity to such features is achieved by our topological losses (see \autoref{fig:wh_good}).
\todo{R2.2c}\markup{Despite neglecting label triples,} both $TP_{i,j=i}$ and $TP_{i,j \ge i}$ adapt segmentation topology, significantly reducing $BE$ when compared with CCA ($p < 10^{-5}$ and $p < 10^{-6}$, respectively).
It is telling, however, that this improvement constitutes a topologically accurate segmentation in only two of forty cases\markup{, and fails to reduce the $HDD$ by as much as CCA (see \autoref{tab:topo_3D})}\todo{3.4e}
Coupled with the deleterious effect of topological post-processing on spatial overlap (see \autoref{tab:dsc_3D}), these \markup{observations help} to characterise the challenge of this task and the limitations of its CNN-based solution.
Degradation in \markup{$gDSC$} is significant: $p < 10^{-6}$ for both $TP_{i,j=i}$ and $TP_{i,j \ge i}$.

Primarily, such degradation occurs where U-Net segmentation includes multiple, probabilistically credible candidates that, to some extent, are consistent with the topological prior.
\todo{R3.2e}\markup{This might explain why the drop in $DSC$ is particularly pronounced for the atria, whose complex and variable presentation restrict the spatial overlap performance of U-Net, compared with their ventricular counterparts.
For example,} \autoref{fig:wh_cc} demonstrates a case in which U-Net localises the left atrium to three distinct regions of the image.
Since each $la$ component has a persistence greater than 0.9, separating true from anomalous components is probabilistically equivocal.
In this context, $TP_{i,j \ge i}$ results in a segmentation with two left atrium components.
Remote from its true location and of significant size, a spurious component inferior and posterior to the apex limits spatial overlap performance.

Such errors are problematic in and of themselves.
However, given its combinatorial nature, they can also confound multi-class topology.
The prior seeks to isolate $la$ from $ra$ by specifying that their combination comprise two connected components.
Implicitly, however, our specification assumes that both structures demonstrate correct topology individually.
Provided a probabilistic segmentation that equivocally suggests multiple atrial components (see \autoref{fig:wh_cc}), CNN optimisation can reduce $L_{topo}$ by reinforcing the communication between atria: a modification at odds with anticipated anatomy.

\begin{table*}[t]
\centering
\caption{Spatial overlap performance on the encrypted MM-WHS test set: multi-class segmentation of high resolution 3D CMR.}
\label{tab:dsc_3D}
\markup{\begin{tabularx}{\textwidth}{*{1}{L}|*{5}{L}|*{2}{L}}
\toprule
&
    \multicolumn{5}{c|}{\textbf{Per class performance} (\%)} & \multicolumn{2}{c}{\textbf{Generalised performance} (\%)} \\
&
    $DSC_{la}$ &
    $DSC_{ra}$ &
    $DSC_{lv}$ &
    $DSC_{rv}$ &
    $DSC_{my}$ &
    $gDSC$ &
    $\Delta gDSC$ \\
\toprule
$U$-$Net$ &
    83.3$_{\scaleto{\text{(79.9,86.8)}}{\iqr}}$ & 81.4$_{\scaleto{\text{(72.6,87.2)}}{\iqr}}$ & 91.6$_{\scaleto{\text{(85.8,93.8)}}{\iqr}}$ & 89.3$_{\scaleto{\text{(80.5,93.1)}}{\iqr}}$ & 79.7$_{\scaleto{\text{(75.6,83.2)}}{\iqr}}$ & 85.1$_{\scaleto{\text{(81.2,87.7)}}{\iqr}}$ &
    --- \\
\midrule
$+CCA$ &
    87.3$_{\scaleto{\text{(84.3,89.2)}}{\iqr}}$ & 87.3$_{\scaleto{\text{(82.1,88.9)}}{\iqr}}$ & 91.8$_{\scaleto{\text{(86.3,94.2)}}{\iqr}}$ & 90.5$_{\scaleto{\text{(81.9,93.2)}}{\iqr}}$ & 79.8$_{\scaleto{\text{(75.6,83.2)}}{\iqr}}$ & 86.5$_{\scaleto{\text{(82.1,89.9)}}{\iqr}}$ & \phantom{1}$+$1.3$_{\scaleto{\text{(0.5,2.5)}}{\iqr}}$ \\
\midrule
$+{TP}_{i, j = i}$ &
    71.9$_{\scaleto{\text{(56.0,81.2)}}{\iqr}}$ & 74.5$_{\scaleto{\text{(58.4,82.2)}}{\iqr}}$ & 82.4$_{\scaleto{\text{(77.0,86.9)}}{\iqr}}$ & 78.0$_{\scaleto{\text{(70.3,85.1)}}{\iqr}}$ & 70.2$_{\scaleto{\text{(63.6,74.0)}}{\iqr}}$ & 75.3$_{\scaleto{\text{(68.4,78.6)}}{\iqr}}$ & $-$10.1$_{\scaleto{\text{(-14.8,-3.6)}}{\iqr}}$ \\
$+{TP}_{i, j \geq i}$
    & 68.4$_{\scaleto{\text{(58.7,81.2)}}{\iqr}}$ & 67.8$_{\scaleto{\text{(55.9,82.2)}}{\iqr}}$ & 84.1$_{\scaleto{\text{(79.0,86.9)}}{\iqr}}$ & 78.8$_{\scaleto{\text{(69.0,86.9)}}{\iqr}}$ & 70.5$_{\scaleto{\text{(64.3,77.0)}}{\iqr}}$ & 75.9$_{\scaleto{\text{(70.6,80.5)}}{\iqr}}$ & \phantom{1}$-$8.9$_{\scaleto{\text{(-13.7,-4.7)}}{\iqr}}$ \\
\bottomrule
\multicolumn{8}{l}{\scriptsize Reporting:
    $DSC_{\{la,ra,lv,rv,my\}}$ $P_{50_{\scaleto{(P_{25}, P_{75})} {\iqr}}}$;
    $gDSC$ $P_{50_{\scaleto{(P_{25}, P_{75})} {\iqr}}}$;
    $\Delta gDSC$ $P_{50_{\scaleto{(P_{25}, P_{75})} {\iqr}}}$.}
\end{tabularx}}\todo{\normalfont R3.2c}
\end{table*}

\section{Discussion}
\label{sec:discussion}

\subsection{Context}

Frequently concerned with the delineation of structured anatomical targets, medical image segmentation often benefits from the incorporation of prior information.
Our work adds to a rich body of research concerned with methods to leverage and make best use of such priors.
Historically, a motivation for their use has been a desire to constrain predicted segmentations to anatomically credible morphology and multi-class configuration.
We interpret these ambitions through the lens of topology, a property that, whilst long acknowledged as critical to anatomical plausibility, has rarely been considered explicitly.
Instead, topology has more often been captured implicitly, within the examples comprising an atlas or statistical shape model, for example.
In contrast, our work exploits PH, an increasingly popular tool from topological data analysis, exposing the topological features of image data.
Accordingly, topology provides both the motivation for, and unlike the majority of previous work, the mathematical basis of our methodology and assessment. 

Most recently, given the CNN-based state of the art, authors have considered means to inject prior information into parameter optimisation.
Popularised by \cite{Oktay2018}, anatomically constrained neural networks are optimised against a compact, latent representation of anatomically plausible segmentations.
However, this approach assumes that the properties which characterise anatomy (morphology, topology etc.) can be implicitly encoded.
Since then, it has been shown that CNNs optimised against such a latent representation can still make predictions with anatomically implausible features \cite{Painchaud2019}.

Compared with this family of methods, our topological loss functions permit optimisation against an explicit topological prior.
This is beneficial to interpretability, making no assumption as to the faithful representation of anatomically relevant features within a learned representation.
However, perhaps its greatest strength is drawn from the abstract quality of the prior information employed.
Those methods based on a learned representation are necessarily biased on training data.
By contrast, topological priors are abstracted from the expert's knowledge of segmentation targets, rather than their appearance within exemplar data.
This enhances the generalisability of our approach, extending its application to the low data setting.
Consider the twenty training cases available in \autoref{subsec:whole_heart}: it is unlikely that an effective latent representation of highly variable, 3D anatomy could be established from such a sample.
Moreover, by decoupling prior information from data, our approach is less susceptible to performance losses in the presence of out of sample test cases, including pathological structural variation.

Finally, we stress that topology is just one component of anatomical plausibility.
In other respects, such as morphology, our approach is complementary to those based on a learned representation of anatomical segmentations.

\subsection{Adaptability to underlying topological losses}
\label{subsec:form_comp}

\todo{R1.1b}\markup{Primarily, the work collectively described in this article and in its preliminary conference paper \cite{Byrne2021} concerns the extension of existing PH-based topological losses to the multi-class setting.
Relying on the persistence of topological features of both individual and combined labels, our framework is adaptable to a range of foundational PH-based loss formulations.
Of those published we apply our extension to one of the family of topological losses set out in Reference \cite{Bruel-Gabrielsson2019}.
Peripheral to our main focus, we prioritise a conceptually straightforward formulation that depends linearly on the persistence of predicted topological features, comparing with the ground truth topology expressed by a Betti number prior.

More generally, those seeking a PH-based loss function are free to leverage the birth and death of features ($\Delta p = p_{\text{birth}} - p_{\text{death}}$) to expose the differences between predicted and ground truth topologies.
In formulating a scalar distance between the two, previous works have made reference to alternate graphical representations of the PH feature space.
Some of these, such as the persistence barcode (referenced in our work) and diagram might appear to invite different loss formulations, as in the respective works of References \cite{Bruel-Gabrielsson2019} and \cite{Hu2019}.

Associated design choices seem to extend to the description of ground truth topology, the former appearing to favour comparison with an associated persistence diagram, the latter, an abstract Betti number prior.
However, where idealised topology is that of a singular, discrete object (or set of objects), descriptions of feature birth and death are superfluous.
Their associated binary (or one-hot encoded) ground truth segmentations taking values in $\{0,1\}$, all features are born and die at $p=1$ and $p=0$, respectively.
Such features are indistinguishable by their persistence.
Hence, in the binary case, a Betti number prior (whether determined analytically by our prior knowledge, or derived from ground truth data) provides a minimal yet complete description of topology.
More abstractly, these arguments express the fact that the machinery of PH is designed to reveal the topological features of \emph{continuous} data via a series of binary snapshots (see \autoref{fig:PH_barcode}); in most supervised CNN segmentation tasks (as in our work and in \cite{Hu2019}), ground truth labels \emph{are} binary.
Exceptions include cases where target topology is expressed within a probabilistic ground truth (perhaps derived from multiple operators), or where segmentation is reframed as a dense regression task (perhaps as prediction of a distance map).

Irrespective of both graphical representation and topological prior formulation, however, PH-based losses seek a notion of distance (of which there are many) in the topological feature space of predicted probabilistic segmentations, defined in $\mathbb{R}^2$ by paired births and deaths.
For example, in Reference \cite{Hu2019}, the $L2$-norm of the PH feature space is used, whereas, our work relies on the $L1$-norm.
This distinction is a design choice of the underlying PH-based loss formulation, one that is independent from, and entirely consistent with both: the graphical representation of topological features within a persistence diagram, barcode or otherwise; and our framework for multi-class topology.
Experimental assessment of these options is outside the scope of our work.}

\subsection{Computational performance}

Compared with our previous work \cite{Byrne2021}, adopting the CubicalRipser algorithm \cite{Kaji2020} and integrating its functionality with PyTorch conferred significant computational gains.
Basing our previous implementation on \cite{Bruel-Gabrielsson2019}, post-processing a single, short axis slice required over six minutes \cite{Byrne2021}.
Here, topological refinement had a mean execution time of only 7.12~s.

In 3D, these gains are not only convenient, but arguably \emph{enable} practicable scientific investigation.
By our previous implementation, computing the barcode for a single MM-WHS volume took over an hour.
Given that our formulation demands the computation of multiple barcodes at each iterative step, such execution times are impractical.
By contrast, in this work, 3D topological post-processing (including all iterations) required a mean of 15.7 minutes.
Practical post-processing times for 3D data permit the specification of topological priors \emph{truly} related to anatomy rather than its appearance in 2D cross-section.
This averts the technical challenge of anticipating the slice-wise topological changes associated with tomographic reconstruction \todo{R3.5b}\markup{(only overcome in \autoref{subsec:short_axis} by manually excluding basal and apical slices)}
and
makes our formulation generalisable to many multi-class segmentation tasks.

\subsection{Limitations}

We have demonstrated our approach within two CMR segmentation tasks, exhibiting varying degrees of success.
Whilst the proposed loss functions reliably improve segmentation topology, they can also be associated with degradation in other metrics of performance when compared against U-Net prediction.
Prior to topological fine tuning, CNN pre-training establishes an optimal set of parameters with respect to spatial overlap.
Hence by introducing topological priors, some compromise is to be expected.
Our experiments demonstrate that the extent to which this trade off is felt depends strongly on the task considered.
In 2D short axis segmentation, topological post-processing coincided with a small increase in spatial overlap performance.
However, in 3D, topological improvement was associated with a statistically significant drop in $gDSC$.
Aside from the limitations of our approach, this result indicates the inability of pixel-wise loss functions to promote topological feature learning.

\begin{figure}[t]
\centering
\includegraphics[width=0.8\linewidth]{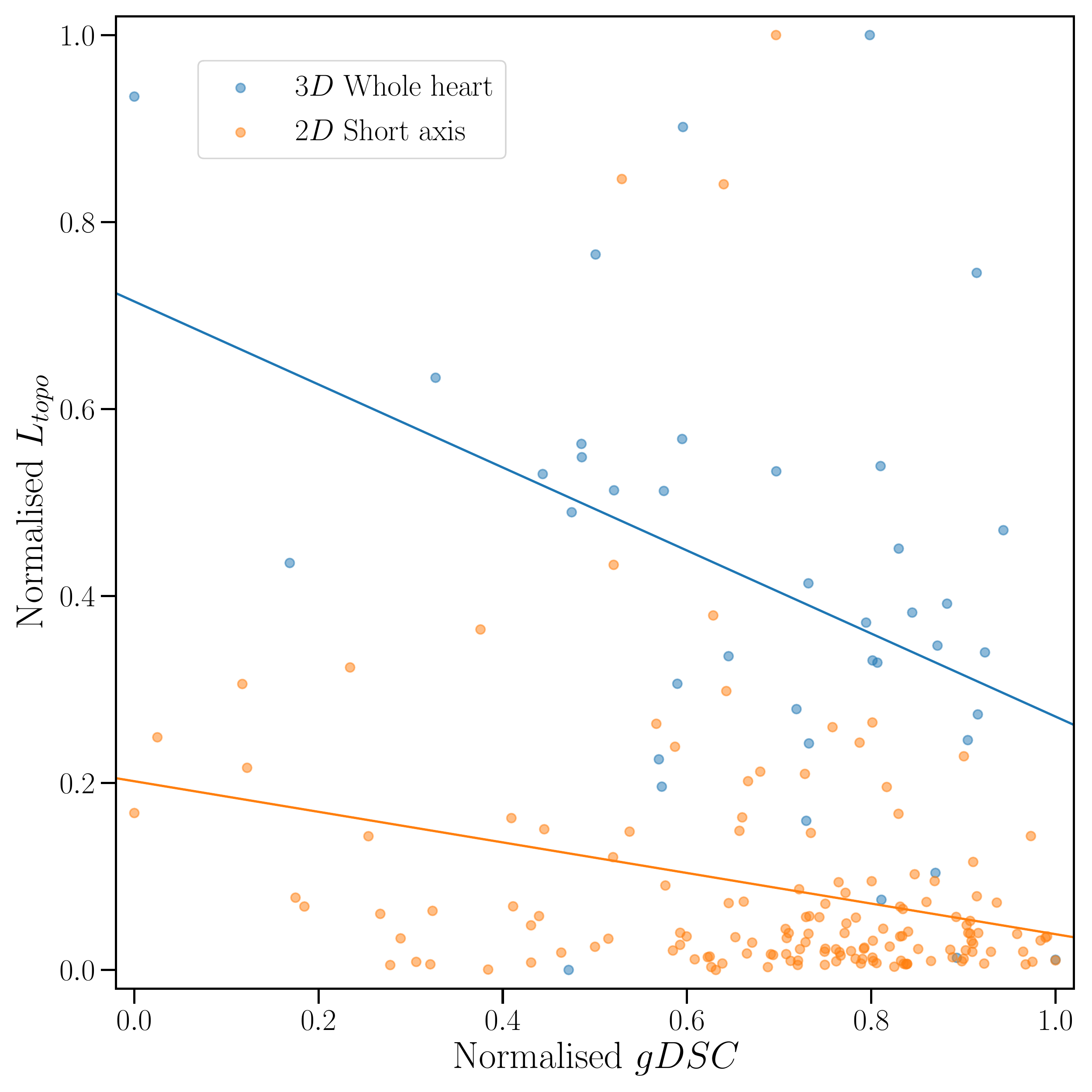}
\caption{The relationship between topological performance and spatial overlap, prior to post-processing.
For inspection, metrics are normalised across their respective test sets.}
\label{fig:correlation}
\end{figure}

We assert that this difference is strongly associated with pre-trained U-Net performance.
A rich body of research indicates the successes of CNN-based short axis segmentation \cite{Chen2020}.
Compared with the majority of 3D applications, this task benefits from numerous training data, reduced structural variability and often a surplus of computational resource in relation to adequate model capacity.
This culminates in vastly improved CNN-based segmentation of the 2D short axis image compared with the 3D whole-heart.
Qualitatively, we suggest that the former task is more closely aligned with our contextual assumption for the application of topological post-processing: that pre-trained CNN segmentation closely approximates the ground truth aside from a small number of topological errors of limited spatial extent.
Quantitatively, we speculate that the applicability of our approach is related to the extent to which pre-trained CNN topological error is explained by spatial overlap performance.
As illustrated in \autoref{fig:correlation}, Spearman's Rho suggests that in the 3D task, there is greater association between U-Net spatial overlap and topological performance (-0.364 versus -0.228).
This suggests that the topological errors presented by U-Net are more likely to be associated with significant deficits in spatial overlap.
Equivocation between true and spurious topological features results.
\todo{R2.9}\markup{Hence, compared with our 2D experiment, 3D topological improvements could only be achieved after reducing $\lambda$, and relaxing the constrained dissimilarity with pretrained U-Net prediction.}
Validation of this assessment across a range of tasks will be necessary to fully understand the generalisability of our approach.

Finally, whilst we remain concerned by degradation in $gDSC$, we stress the importance of anatomically meaningful segmentation to an array of downstream tasks, including surgical planning \cite{Valverde2017}.
For these purposes, we think it important that diverse aspects of performance, including topology, are represented and considered alongside spatial overlap.

\subsection{Future work}

Our formulation presents a general framework that is applicable to any multi-class image segmentation task.
In particular, we are keen to explore its application to problems in which anatomical topology varies from case to case, possibly indicating clinically relevant structural disease.
A patient-specific topological prior lends itself to the treatment of congenitally malformed cardiac anatomy, for example.

More generally, we speculate that topological loss functions may enhance a range of CNN training paradigms.
Given our efficient implementation, our approach could be incorporated into conventional CNN optimisation.
Moreover, the abstract and explicit priors on which such losses are based could plausibly provide a supervisory signal for weakly or semi-supervised learning.

\section{Conclusion}
\label{sec:conclusion}

We have extended PH-based loss functions to multi-class image segmentation.
In the context of state of the art spatial overlap performance, our novel approach made statistically significant and predictable improvements in label map topology within 2D and 3D tasks.
Our approach is theoretically founded, building a multi-class prior as the collection of individual and paired label map topologies.
Compared with the naive consideration of singular labels (the natural extension of previous work), the superiority of this scheme is borne out experimentally.
Crucially, we adopted a highly efficient algorithmic backbone for the computation of cubical PH, achieving dramatic improvements in execution time and permitting practicable extension to the 3D setting.
A careful analysis of both quantitative and qualitative performance allowed us to reflect the limitations of our approach and consider its wider application.
Whilst demonstrated in the field of CMR image analysis, our formulation is generalisable to any multi-class segmentation task: we envisage many applications across a diverse range of anatomical and pathological targets.


\section{Appendix\\Performance by complex construction}
\label{sec:appendix}

\todo{NB.1b}\markup{\autoref{tab:construction} extends the results achieved in Experiment 1 (segmentation of 2D short axis data) to consider the performance of multi-class topological losses when relying on the different modes of cubical complex construction described in \autoref{subsec:hom_cub_comp}.
We use superscripts to differentiate: $TP_{\cdot, \cdot}^{0}$ and $TP_{\cdot, \cdot}^{2}$, reflecting topological post-processing in conjunction with cubical complexes in which pixels are considered 0- and 2-\emph{cells}, respectively.
Likewise, and as demonstrated in \autoref{fig:cub_comp}(a) and (b), we refer to these as the 0- and 2-\emph{construction}.
Investigation was conducted as described in \autoref{subsec:short_axis}, including the definition of the generalised Betti Error ($BE$) and Topological Success ($TS$) metrics defined in \autoref{subsec:metrics}.
Compared with the per class metrics reported in \autoref{tab:topo_2D} and \autoref{tab:dsc_2D}, reflecting $TP_{.,.}^0$, equivalent results for $TP_{.,.}^2$ were unremarkable: hence for brevity, we omit $DSC$ and $HDD$ results from \autoref{tab:construction}.

In the context of post-processing according to individual class labels only ($i, j=i$), rows one and two compare the performance of each mode of complex construction.
Whilst these suggest no difference in the median $BE$ (1.00 in both cases), marginal improvement in $TS$ (98.0 - 97.3) is associated with the 2-\emph{construction}.
This pattern is largely replicated in rows three and four, which reflect the application of topological priors for class labels and their pairwise combination ($i,j \ge i$).
In these rows, the 2-\emph{construction} is associated with small improvement in $BE$ (0.00 - 0.51).
This translates into a marginal gain in $TS$ (99.3 - 98.7).

However, irrespective of the choice of topological prior, there are no statistically significant differences in topological (or spatial overlap) performance between either the 0- or 2-\emph{construction}.
In the absence of any appreciable difference, we employ the 0-\emph{construction} throughout this work (the default within the CubicalRipser package), and the account provided in the main text.
We anticipate that broadly, our findings were not affected by this choice.}

\renewcommand{\iqr}{4.5pt}
\begin{table}[h]
\centering
\caption{Extension to Experiment 1:\\Dependence on mode of cubical complex construction.}
\label{tab:construction}
\scriptsize
\markup{\begin{tabularx}{\columnwidth}{ *{1}{L} | *{2}{L} | *{2}{L} }
\toprule
&
    \multicolumn{2}{c|}{\textbf{Spatial overlap performance (\%)}} & \multicolumn{2}{c}{\textbf{Topological performance}} \\
&
    $gDSC$ &
    $\Delta gDSC$
    & $BE$
    & $TS (\%)$ \\
\toprule
${TP}_{i, j = i}^{0}$ &
    93.4$_{\scaleto{\text{(91.6,94.3)}} {\iqr}}$ &
    $+$0.0$_{\scaleto{\text{(0.0,0.1)}} {\iqr}}$ &
    1.00$_{\scaleto{\text{(1.00,2.00)}} {\iqr}}$ &
    97.3$_{\scaleto{\text{(16.2)}} {\iqr}}$ \\
${TP}_{i, j = i}^{2}$ &
    93.3$_{\scaleto{\text{(91.7,94.3)}} {\iqr}}$ &
    $+$0.0$_{\scaleto{\text{(0.0,0.1)}} {\iqr}}$ &
    1.00$_{\scaleto{\text{(0.00,1.00)}} {\iqr}}$ &
    98.0$_{\scaleto{\text{(14.0)}} {\iqr}}$ \\
\midrule
${TP}_{i, j \geq i}^{0}$
    & 93.3$_{\scaleto{\text{(91.7,94.3)}} {\iqr}}$ & $+$0.0$_{\scaleto{\text{(0.0,0.1)}} {\iqr}}$ &
    0.51$_{\scaleto{\text{(0.00,1.00)}} {\iqr}}$ &
    98.7$_{\scaleto{\text{(11.3)}} {\iqr}}$ \\
${TP}_{i, j \geq i}^{2}$ &
    93.3$_{\scaleto{\text{(91.7,94.4)}} {\iqr}}$ &
    $+$0.0$_{\scaleto{\text{(0.0,0.1)}} {\iqr}}$ &
    0.00$_{\scaleto{\text{(0.00,1.00)}} {\iqr}}$ &
    99.3$_{\scaleto{\text{(8.3)}} {\iqr}}$ \\
\bottomrule
\multicolumn{5}{l}{\tiny Reporting:
    $gDSC$ $P_{50_{\scaleto{(P_{25}, P_{75})} {\iqr}}}$;
    $\Delta gDSC$ $P_{50_{\scaleto{(P_{25}, P_{75})} {\iqr}}}$;
    $BE$ $P_{99_{\scaleto{(P_{98}, P_{100})}{\iqr}}}$;
    $TS$ $\rho_{\scaleto{(\sigma_{\rho})}{\iqr}}$.}
\end{tabularx}}
\end{table}

\section{Code availability}
\label{sec:code_availability}

The source code for our implementation of multi-class topological losses is available at: \url{https://github.com/nick-byrne/topological-losses.git}.

\end{document}